\documentclass{emulateapj}

\slugcomment{}
\usepackage{graphicx}
\usepackage{longtable}

\shorttitle{How Do Galaxies Assemble Their Mass At $Z>3$?}
\shortauthors{Lee et al.}
\begin{document}
\def\hh{\, h^{-1}}
\newcommand{\ie}{$i.e.$,}
\newcommand{\wth}{$w(\theta)$}
\newcommand{\mpc}{Mpc}
\newcommand{\xir}{$\xi(r)$}
\newcommand{\Lya}{Ly$\alpha$}
\newcommand{\Lyb}{Lyman~$\beta$}
\newcommand{\Hb}{H$\beta$}
\newcommand{\msun}{M$_{\odot}$}
\newcommand{\hmsun}{$h^{-1}$M$_{\odot}$}
\newcommand{\sfr}{M$_{\odot}$ yr$^{-1}$}
\newcommand{\dnsty}{$h^{-3}$Mpc$^3$}
\newcommand{\za}{$z_{\rm abs}$}
\newcommand{\ze}{$z_{\rm em}$}
\newcommand{\cmtwo}{cm$^{-2}$}
\newcommand{\nhi}{$N$(H$^0$)}
\newcommand{\degpoint}{\mbox{$^\circ\mskip-7.0mu.\,$}}
\newcommand{\halpha}{\mbox{H$\alpha$}}
\newcommand{\hbeta}{\mbox{H$\beta$}}
\newcommand{\hgamma}{\mbox{H$\gamma$}}
\newcommand{\kms}{\,km~s$^{-1}$}      
\newcommand{\minpoint}{\mbox{$'\mskip-4.7mu.\mskip0.8mu$}}
\newcommand{\mv}{\mbox{$m_{_V}$}}
\newcommand{\Mv}{\mbox{$M_{_V}$}}
\newcommand{\peryr}{\mbox{$\>\rm yr^{-1}$}}
\newcommand{\secpoint}{\mbox{$''\mskip-7.6mu.\,$}}
\newcommand{\sqdeg}{\mbox{${\rm deg}^2$}}
\newcommand{\squig}{\sim\!\!}
\newcommand{\subsun}{\mbox{$_{\twelvesy\odot}$}}
\newcommand{\et}{{\it et al.}~}
\newcommand{\er}[2]{$_{-#1}^{+#2}$}
\def\h50{\, h_{50}^{-1}}
\def\hbl{km~s$^{-1}$~Mpc$^{-1}$}
\def\ltsima{$\; \buildrel < \over \sim \;$}
\def\simlt{\lower.5ex\hbox{\ltsima}}
\def\gtsima{$\; \buildrel > \over \sim \;$}
\def\simgt{\lower.5ex\hbox{\gtsima}} 
\def\arcs{$''~$}
\def\arcm{$'~$}
\newcommand{\wu}{$U$}
\newcommand{\wb}{$B_{435}$}
\newcommand{\wv}{$V_{606}$}
\newcommand{\wi}{$i_{775}$}
\newcommand{\wz}{$z_{850}$}
\newcommand{\hmpc}{$h^{-1}$Mpc}
\newcommand{\lm}{$L$--$M$}
\newcommand{\ws}{$\mathcal{S}$}
\newcommand{\wm}{$\mathcal{M}$}
\newcommand{\sm}{$\mathcal{S}$-$\mathcal{M}$}
\newcommand{\medianLM}{$\tilde{\mathcal{L}}(M)$}
\newcommand{\mf}{$\phi_\mathcal{M}$}

\title{How Do Star-forming Galaxies at $z>3$ Assemble Their Masses?}
\author{Kyoung-Soo Lee\altaffilmark{1}, Henry C.~Ferguson\altaffilmark{2}, Tommy Wiklind\altaffilmark{3}, Tomas Dahlen\altaffilmark{2}, Mark E. Dickinson\altaffilmark{4}, \\Mauro Giavalisco\altaffilmark{5}, Norman Grogin\altaffilmark{2}, Casey Papovich\altaffilmark{6}, Hugo Messias\altaffilmark{7},  Yicheng Guo\altaffilmark{5}, Lihwai Lin\altaffilmark{8}}
\altaffiltext{1}{Department of Physics, Purdue University, West Lafayette, IN 47907}
\altaffiltext{2}{Space Telescope Science Institute, 3700 San Martin Drive, Baltimore, MD 21218}
\altaffiltext{3}{Joint ALMA Observatory, Santiago, Chile}
\altaffiltext{4}{National Optical Astronomy Observatory, Tucson, AZ 85719}
\altaffiltext{5}{Department of Astronomy, University of Massachusetts, Amherst, MA 01003}
\altaffiltext{6}{Department of Physics and Astronomy, Texas A\&M University, College Station, TX 77843}
\altaffiltext{7}{Departamento de Astronom\'ia, Universidad de Concepci\'on, Chile}
\altaffiltext{8}{Institute of Astronomy \& Astrophysics, Academia Sinica, Taipei 106, Taiwan}
\begin{abstract}
We investigate how star-forming galaxies typically assemble their masses at high redshift. Taking advantage of the deep multi-wavelength coverage of the GOODS data set, we select two of the largest samples of high-redshift star-forming galaxies based on their UV colors  and measure stellar mass of individual galaxies. We use template-fitting photometry to obtain optimal estimates of the fluxes in lower-resolution ground-based and {\it Spitzer} images using prior information about galaxy positions, shapes, and orientations. By combining the data and realistic simulations to understand measurement errors and biases, we make a statistically robust determination of stellar mass function (SMF) of the UV-selected star-forming galaxies at $z\sim4$ and $5$. We report a broad correlation between stellar mass and UV luminosity, such that more UV-luminous galaxies are, on average, also more massive. However, we show that the correlation has a substantial intrinsic scatter, particularly for UV-faint galaxies, evidenced by the fact there is a non-negligible number of UV-faint but massive galaxies.   Furthermore, we  find that the low-mass end of the SMF does not rise as steeply as the UV luminosity function ($\alpha_{\rm{UVLF}}\approx-(1.7-1.8)$ while $\alpha_{\rm{SMF}}\approx-(1.3-1.4)$) of the same galaxies.  In a smooth and continuous formation scenario where star formation rates are sustained at the observed rates for a long time, these galaxies would have accumulated more stellar mass (by a factor of $\approx3$) than observed and therefore the stellar mass function would mirror more closely that of the UV luminosity function. The relatively shallow slope of the SMF is due to the fact that many of the UV-selected galaxies   are not massive enough, and therefore are too faint in their rest-frame optical bands, to be detected in the current observations. Our results favor a more episodic formation history in which star formation rates of low-mass galaxies  vary significantly over cosmic time, a scenario currently favored by galaxy clustering. Our findings for the UV-faint galaxies at high redshift are in contrast with previous studies on more UV-luminous galaxies, which exhibit a tighter SFR-$M_{\rm{star}}$ correlation. The discrepancy may suggest that galaxies at different luminosities may have different evolutionary paths. Such a scenario presents a nontrivial test to theoretical models of galaxy formation.
\end{abstract}
  

\section{Introduction}

Our knowledge of the distant universe has expanded substantially over the last decade thanks to deep multi-wavelength surveys. Currently, galaxies are routinely identified out to $z\sim7$ and beyond, and their key quantities such as star formation rates (SFR), stellar masses ($M_{\rm{star}}$), and extinction are measured via spectroscopy and multi-wavelength photometry \citep[e.g,][]{papovich01,shapley01, stark09,mannucci09}. Large-area surveys have also enabled us to characterize the statistical properties of the high-redshift galaxy population as a whole. The UV luminosity function primarily reflects the distribution of star formation rates within the population and cosmic star formation rate density \citep{steidel99,mauro04a,ouchi04a,sawicki05, bouwens07,reddy09,mclure09, bouwens11a}. The stellar mass function traces the distribution of stellar mass within the population and cosmic stellar mass density \citep{drory05,fontana06,marchesini09,marchesini10,gonzalez11}. Morphological studies have quantified the relative spatial distribution and size of stellar components \citep{ferguson04, bouwens04c, ravindranath06,lotz06}. Finally, the galaxy clustering measures the large-scale spatial correlation between galaxies and how it compares with that of underlying dark matter and dark matter halos \citep{GD01,ouchi04b,ouchi05,adelberger05,quadri07,lee06,lee09}.  Careful comparison of these statistical quantities at different cosmic epochs can provide powerful tools to gain  insight into the evolutionary sequence of galaxy assembly in the high-redshift universe. 

This tremendous observational progress offers the opportunity to establish whether galaxy growth is dominated by smooth accretion or more episodic events. While both modes certainly exist and are observed at high redshift,  the relative importance of the two modes is currently unconstrained at $z>3$ \citep[but see][for measurements out to $z\sim2$]{rodighiero11,elbaz11}. Furthermore, there are observational evidence that the dominant mode may vary with galaxy or halo mass \citep[e.g.,][]{vanderwel11}.  A key measurement that directly constrains the main mode of star formation, or average star formation history, is the location of galaxies on the SFR-$M_{\rm{star}}$ plane. If galaxies assemble smoothly over a time scale comparable to the Hubble time, the majority of galaxies would form a tight sequence with a slope of close to unity \citep{noeske07}. On the other hand, in the case of bursty/episodic star formation, one should observe significant scatter about the mean relation. 

Unfortunately, at  high redshift, different studies report conflicting results. Some find a strong correlation with a tight scatter \citep{daddi07a,pannella09,magdis10,lee11,sawicki11} while others find no or weak correlation with larger scatter \citep[][]{shapleyetal05,reddy06a,mannucci09}.  Some of this discrepancy may be real, if the main mode of galaxy growth changes with stellar mass or host halo mass \citep[][]{renzini09,lee09}, or with different ``types'' of galaxies. However, it is likely that the discord between different measurements may be at least in part due to different and poorly understood selection effects suffered by different samples and/or different calibration methods employed by these studies to derive various star formation rates and stellar mass. 

Some of the existing samples of galaxies are based on a UV color-selection while some others use a combination of UV and optical colors. Some even require an additional detection in the mid-infrared (such as {\it Spitzer}) complicating the interpretations of their results in the global picture. Calibration is another issue at hand. For example, SFRs derived from polycyclic aromatic hydrocarbon (PAH) emission are known to depend strongly on galaxy metallicity and the internal distribution of HII regions and dust \citep{calzetti07} even when the precise redshift is known. In addition, the presence of active galactic nuclei (AGN) in some of high-redshift galaxies may contribute to the observed flux in the mid-infrared \citep[e.g.,][]{daddi07b}, and subsequently add to the observed scatter in the SFR-$M_{\rm{star}}$ relation unless such galaxies are identified or the AGN contribution is accounted for. These complications make it challenging to constrain the intrinsic SFR-$M_{\rm{star}}$ scaling law even for galaxies with the most comprehensive multiwavelength and spectroscopic coverage. Furthermore, at $z>3$, even these diagnostics become out of reach as the PAH emission redshifts to the longward of 24~$\mu$m \citep[and H$\alpha$ redshifts to $>3$$\mu$m; but see][]{shim11}, and only galaxies with the most extreme starbursts such as sub-mm galaxies can be detected via other {\it Spitzer} MIPS bands or {\it Herschel}. 

Here we take an alternative approach to constrain the main mode of star formation by using two statistical measurements of galaxies, namely, the UV luminosity function (UVLF) and stellar mass function (SMF) of the {\it same} galaxies. The former constrains the distribution of  {\it recent} star formation in galaxies, while the latter measures the distribution of existing stellar mass in the same galaxies directly tracing the integral of all past SF activities. Hence, these two quantities represent the statistical distribution of the SFR (modulo dust) and stellar mass rather than those in individual galaxies. In essence, comparison of the two distributions will shed light on how one traces the other (when the effect of dust extinction is accounted for). If they trace each other closely, the overall shape of the two distributions will be similar. Alternatively, any difference between the two distributions can be attributed to the statistical mapping between the two as the two quantities are measured from the same galaxy population. 

There are several advantages of this approach over the direct method discussed previously. First, as we will show later, the two quantities, UVLF and SMF, can be determined very robustly by combining the observations and extensive simulations via rigorous statistical analyses. Second, our approach bypasses the need to measure  star formation rates very accurately for individual galaxies. As mentioned previously, the estimation of star formation rates is hard even when precise photometry  and redshift information are available, and has been the primary challenge to make progress. Instead, we take advantage of the deepest surveys to statistically map the UV luminosity to SFRs \citep{bouwens09, reddy10,castellano11,finkelstein11}. In particular, at $z>3$, this approach offers new possibilities to obtain useful insights into the typical mode of galaxy growth as deep imaging surveys are being conducted in the near- and mid-infrared from space \citep[e.g., {\it Spitzer} Extended Deep Survey, Cosmic Assembly Near-infrared Deep Extragalactic Legacy Survey; ][]{grogin11,koekemoer11}. 
 
 In this paper, we investigate how star-forming galaxies assemble their mass in the first two billion years. We take advantage of two of the deepest survey fields with the most comprehensive multi-wavelength coverage, namely, GOODS-N and GOODS-S field. The combination of the large areal coverage ($\approx$0.1 degree$^2$ total) and depth  achieved in these fields allows us to identify a large number of high-redshift star-forming galaxies and robustly measure their population properties from integrated galaxy light. We use the TFIT template-fitting photometry package to carry out the photometry and present the most extensive set of simulations to date to validate the techinque and quantify the uncertainties. As we will demonstrate later, TFIT greatly improves the photometric accuracy for individual measurements over the conventional methods as well as dramatically increases the number of sources with reliable multi-wavelength photometry\footnote{For example, \citet{stark09} noted that only 35\% of the sources at $z\sim4$ are sufficiently isolated to perform robust aperture photometry. On the other hand, this work using TFIT rejects less than 5\% due to blending.}.

Throughout this work, we use $(\Omega_m, \Omega_\Lambda, \sigma_8, h_{100})=(0.28, 0.72, 0.9, 0.72)$. Magnitudes are given in AB system \citep{oke83} unless noted otherwise.

\section{Data, Sample Selection, and Multi-wavelength Photometry}
\subsection{Data}
The data set we use for our analyses consists of those obtained as part of the Great Observatories Origins Deep Survey \citep[GOODS; ][]{mauro04a}. The two fields, GOODS-N and GOODS-S, each covers roughly 160 arcmin$^2$, include deep $HST$/ACS $B_{435}V_{606}i_{775}z_{850}$ (F435W, F606W, F775W, and F850LP) and deep {\it Spitzer} IRAC [3.6$\mu$m], [4.5$\mu$m], [5.8$\mu$m]\footnote{We note that the majority of our samples are not detected in the [5.8$\mu$m] and [8.0$\mu$m] bands} and MIPS [24$\mu$m] data. The GOODS-S field is covered by deep $J$, $H$, and $K_S$-band data taken with the ISAAC camera,  and the $U$-band taken with the Visible Multiobject Spectrograph (VIMOS)  on the Very Large Telescope \citep[VLT: ][respectively]{retzlaff10,nonino09}. The GOODS-N field is covered by the $J$ and $K_S$-band data taken with the Wide-Field Infrared Camera (WIRCAM) on the Canada-France-Hawaii Telescope \citep[CFHT;][L. Lin et al., in prep]{wang10}, and the $U$-band data taken  with the MOSAIC prime focus camera \citep{capak04} on the Kitt Peak National Observatory 4-m telescope. 

\begin{table*}
\begin{center}
\caption{The $5\sigma$ Limiting Magnitudes of the Data Set\label{tbl_1}}
\begin{tabular}{crrrrrrrrrrr}
\\
\tableline\tableline
Field & $U$ & $B_{435}$ & $V_{606}$ & $i_{775}$ & $z_{850}$ & $J$ & $H$ & $K_S$ & [3.6$\mu$m] & [4.5$\mu$m] & [5.8$\mu$m]\\
\tableline
GOODS--N& 27.2 & 28.4 & 28.6 & 27.9 & 27.6 & 25.0& --- & 24.5 & 26.2& 25.6 & 23.5 \\
GOODS--S& 28.1 & 28.4 & 28.6 & 27.9 & 27.6 & 25.7 & 25.5 & 25.1 & 26.2& 25.6& 23.5\\
\tableline
\end{tabular}
\tablecomments{For the ACS bands, the limiting magnitudes are computed within an 0\farcs2 diameter aperture. For the rest, isophotal aperture defined in the detection band is used (see \S\ref{tfit}) }
\end{center}
\end{table*}

The $5\sigma$ limiting magnitudes of the data are tabulated in Table \ref{tbl_1}. Therein, the limiting magnitudes for the ACS bands are computed within an 0\farcs2 diameter aperture. For the rest of the photometric bands, we compute the photometric uncertainties ($1\sigma$ errors) within the isophotal apertures defined in the detection band ($z_{850}$-band; see \S\ref{tfit} for details).

\subsection{``Dropout'' Samples at $z\sim4$ and $5$}
We adopt the Lyman-break technique \citep{steidel96} to identify high-redshift star-forming galaxies (``Lyman Break Galaxies'' or LBGs, hereafter) at $z\sim4$ and $\sim5$ with the color selection criteria discussed in  \citet{mauro04b} and \citet{lee06,lee09}. Extensive spectroscopy campaigns have shown that these selection methods are very robust with minimal contamination by interlopers \citep{vanzella06,vanzella09}. While the adopted selection criteria differ slightly from others found in the literature \citep[e.g.,][]{bouwens07}, our  main conclusions should remain identical to those derived using different selection as  the proper account of the selection efficiency will make appropriate corrections to compensate for the difference.  We used the UV color criteria used by \citet{mauro04b} to select  galaxies at $z\sim4$ and $z\sim5$ as
\begin{eqnarray}\label{color_selection_B}
(B_{435} - V_{606}) &\geq& 1.2 + 1.4 \times (V_{606}-z_{850}) \wedge  \nonumber \\
(B_{435} - V_{606}) &\geq& 1.2 \wedge V_{606}-z_{850} \leq 1.2,  \nonumber
\end{eqnarray}
and
\begin{eqnarray}\label{color_selection_V}
(V_{606}-i_{775}) &>& 1.5+0.9 \times (i_{775}-z_{850}) \vee  \nonumber \\
(V_{606}-i_{775}) &>& 2.0 \wedge (V_{606}-i_{775}) \geq 1.2 \wedge  \nonumber \\ 
(i_{775}-z_{850}) &\leq& 1.3 \wedge \nonumber \\
{\rm S/N}(B_{435})&<&2,\nonumber
\end{eqnarray}
where the symbols $\vee$ and $\wedge$ are the logical ``OR'' and ``AND'' operators, respectively. In both samples, only sources with ${\rm S/N}(z_{850})\geq6$ are considered. In addition to the color criteria, we also removed sources that have stellarity index \citep[using the CLASS\_STAR parameter in the SExtractor;][]{bertina96} greater than 0.8  from the sample when the source is brighter than 26.2 mag in the detection band ($z_{850}$-band). For sources fainter than this limit, our simulation suggests that the stellarity measurement is not as reliable. 

The number of photometric candidates satisfying these selection criteria is 3088 and 987 at $z\sim4$ and 5, respectively. We refer to these samples as $B_{435}$-band and $V_{606}$-band dropouts hereafter as their targeted redshift range require them to ``drop out'' in those passbands. Finally, we removed the most likely low-redshift interlopers based on photometric redshifts from \citet{dahlen10}, which are calibrated against the largest compilation of spectroscopic redshifts available in the GOODS fields \citep[][Stern et al., in prep]{wirth04,cowie04,vanzella06,vanzella09}.  Using their estimates, we remove sources whose total integrated probability $P(z<3)$ is greater than 70\%.  The number of sources removed using this criterion is 136 and 131 galaxies at $z\sim4$ and $5$, respectively. Figure 15 and 18 of \citet{dahlen10} show the distribution of photometric redshifts for the $B_{435}-$ and $V_{606}$-band dropout samples. 

\subsection{{\it Spitzer} MIPS-detected Sources}\label{mips_sources}
We identify sources with the {\it Spitzer} MIPS 24~$\mu$m detection (with the formal signal-to-noise ratio S/N$\ge3$). A bona-fide high-redshift galaxy at $z>3$ with 24~$\mu$m detection (sampling the rest-frame $\lambda\approx 4 - 6~ \mu$m) implies the presence of unusually strong hot dust components produced by either an AGN or an extreme starburst (SFR$\gtrsim$ several hundreds $M_\odot$yr$^{-1}$). Indeed, of the $B_{435}$- and $V_{606}$-band dropouts that are genuinely detected at 24~$\mu$m, several are detected in the deep X-ray data \citep{alexander03,luo08}, further supporting the possibility that they harbor an AGN. We cross-matched the GOODS MIPS catalogs with the dropout lists using a tolerance of 0\farcs5 to minimize blending problems. For the $B_{435}$-band dropouts in the GOODS-N, we find 24 sources with MIPS detections (four of which are detected in X-ray). Of those, 11 objects have spectroscopic redshifts (3 are foreground interlopers at $z<3$ and the other 8 at $3.403<z<4.604$). In the GOODS-S, we find 24 MIPS detection (five of which are X-ray detected). Of the total of 7 objects with spectroscopic redshifts, one lies at $z=2.797$ while the remaining 6 lie at $3.055<z<3.891$ on the low side of the redshift distribution for the $B_{435}$-band dropouts. Four of the six sources at  $3.055<z<3.891$  have X-ray detections. In summary, 44 out of 48 MIPS detections in the combined GOODS-North and South field have $f(24~\mu m)<82~ \mu$Jy. Three spectroscopically confirmed $B_{435}$-band dropouts have $100~ \mu$Jy$<f(24~\mu m)<$$160~\mu$Jy and are also X-ray sources, thus likely AGN. 

As for the $V_{606}$-band dropouts in the GOODS-N, we find 4 MIPS detections none of which has spectroscopic redshifts available. However, two of them are very bright in 24~$\mu$m (113~$\mu$Jy and 347~$\mu$Jy), and thus are very likely to be foreground interlopers. In the GOODS-S, we find 11 MIPS detections three of which are confirmed to be foreground interlopers ($z$=1.324, 1.981, and 3.513) while another lies at $z=4.762$ on the low-side of the redshift distribution for the $V_{606}$-band dropouts. The galaxy at $z=4.762$ is also a known submillimeter galaxy \citep{coppin09}. Of those without spectroscopic redshifts, one is unusually bright (109~$\mu$Jy) to be at $z>4.5$. In summary, for the combined GOODS-N and GOODS-S $V_{606}$-band dropouts, 12 out of 15 MIPS detections have $f(24~\mu m)<60~\mu$Jy. Two with no spectroscopic confirmation have $f(24~\mu m)\gtrsim110~\mu$Jy (no X-ray detection) and another has $f(24~\mu m)=347~\mu$Jy (also without spec-$z$ or X-ray). 

Based on these statistics, we conclude that rather few bona fide $B_{435}$-band dropouts, and almost no $V_{606}$-band dropouts are detected at $24~\mu$m. Thus, we exclude all the sources with MIPS detection from our sample. However, we note that inclusion or exclusion of these sources makes little difference in our main conclusions as their number is few. 

\subsection{Template Fitting Photometry (TFIT)}\label{tfit}
Reliable estimates of physical parameters begin with reliable photometry. This is a challenge for mixed-resolution datasets like GOODS, where the PSF width varies by an order of magnitude across the full wavelength range. Galaxies that are cleanly identified in the HST images may be horribly blended with their neighbors in the {\it Spitzer} images. 

Template Fitting (TFIT) photometry is designed to overcome these challenges and provide accurate flux estimates and colors for galaxies in multi-wavelength mixed-resolution data set such as the data used in this work. While we refer interested readers to  \citet{papovich01} and \citet{laidler07} for more details of the TFIT algorithm, we also note that similar algorithms have been implemented by others to obtain optimal photometry \citep[e.g.,][]{fernandezsoto99,labbe06,grazian06,desantis07}. Here, we introduce basic concepts of TFIT to highlight the main difference between this work and previous ones in the literature. We further demonstrate the effectiveness of TFIT for faint high-redshift galaxies by directly comparing the TFIT results with the conventional methods, namely, aperture photometry. \\

The basic premise of TFIT is that the best flux of a source can be estimated by simultaneous fitting of all the adjacent sources rather than just performing photometry within a fixed aperture. This strategy is particularly effective for crowded fields where sources are often substantially blended with one another and PSFs are large enough to not just affect their immediate neighbors but possibly those further out from any given source. Even though deep fields such as GOODS are by design mostly devoid of bright sources in the optical bands, the same field is crowded in the deep IRAC data. 

In order to achieve simultaneous fitting of multiple sources, TFIT constructs a realistic model of how an object would appear in a low-resolution data based on the observed morphology of the object in a deep high-resolution image. The cutout image of each object is created from the high-resolution data, then convolved with the kernel designed to reproduce the PSF of the low resolution image and block-averaged to the desired pixel scale. The image created with this procedure is called a {\it template}, which provides the best approximation of how the same object appears in the low-resolution image. This assumes no significant dependence of galaxy morphology on wavelength -- i.e., no morphological $k$-correction. While such an assumption may not be valid in some cases, this can be in part overcome by using the high-resolution image observed in the wavelength range closest to the low-resolution one. For the GOODS TFIT photometry, we use the $z_{850}$ band to create templates for IRAC photometry and the $B_{435}$ band for the $U$-band photometry. 

\begin{figure*}[t]
\epsscale{1.0}
\plotone{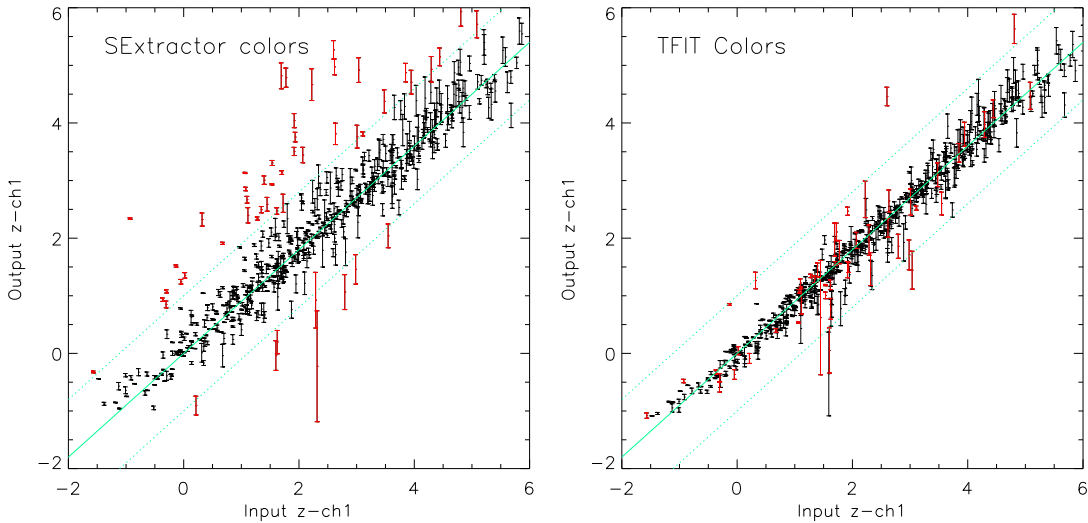}
\caption[sim_SED]{Optimal multiwavelength photometry by TFIT.  Extended sources with a range of colors are simulated to compare the performance of conventional aperture photometry and TFIT. ({\it Left}) The $z-ch1$ (ACS $z_{850}$-band - {\it Spitzer} [3.6$\mu$m]) colors measured by SExtractor are compared with the input colors for 600 simulated galaxies. Large scatter is expected due to source blending and confusion. ({\it Right}) The colors of the same sources measured by TFIT.  The sources with the ``aperture color'' bias $>$1 mag  are shown in red on both panels. Note that most of the same sources are successfully recovered in TFIT-derived colors. }
\label{tfit_sim}
\end{figure*}

Once the templates of all galaxies are created, TFIT performs chi-square minimizations on subsections of the data, to find the best-fit fluxes for each object. Furthermore, TFIT at this stage corrects for any second-order astrometric misregistration which may be present for data sets taken with different instruments. Such misregistration can bias the measured flux and colors not only for TFIT but also for aperture photometry as well. 

We carried out simulations to compare the performance of TFIT in direct comparisons with aperture photometry returned by SExtractor \citep{bertina96}. Artificial galaxies (with known input fluxes) are inserted into the real data (ACS and IRAC) and fluxes are recovered by both aperture photometry and TFIT. As for aperture photometry, we adopt MAG\_AUTO for both ACS and IRAC as the best-estimate.   Figure \ref{tfit_sim} shows the $z-[3.6\mu$m] colors measured from SExtractor ({\it left}) and TFIT ({\it right}) in comparison with the true input colors of the same sources. It is evident that the colors measured by TFIT are much closer to the true values than those by aperture photometry, and also that photometric scatter is considerably smaller. Furthermore, catastrophic outliers identified in aperture photometry (shown in red in both panels) are minimized in TFIT photometry as they are located much closer to the one-to-one line (green) in TFIT colors. This is not surprising because such errors typically result from significant contamination due to crowding. Such catastrophic failures are expected to increase in deeper images where the source density is higher. We also note that such cases are expected to be frequent for faint high-redshift sources whose flux can be significantly affected by usually-brighter neighbors when the conventional method is adopted. \\

Throughout this work, we utilize the GOODS Grand Unified TFIT catalog\footnote{The catalog is soon to be made available at {\tt http://www.stsci.edu/science/goods/DataProducts/}} to measure the spectral properties of LBGs. The catalog includes the SExtractor photometry of the $HST$/ACS band ($B_{435}$, $V_{606}$, $i_{775}$, $z_{850}$) and TFIT photometry for the remainder of our photometric bands ($UJHK_S$[3.6$\mu$m][4.5$\mu$m][5.8$\mu$m]). The ACS $z_{850}$-band (F850LP) was used as the detection band for the source extraction and to define isophotal area for other ACS bands, and also as high-resolution templates for the TFIT photometry. This ensures that the measured flux either by SExtractor or TFIT will return accurate colors within the same isophotal apertures, and that aperture correction is straightforward to handle. Galaxy colors are computed using the MAG\_ISO for the ACS bands and TFIT fluxes for the others, while MAG\_AUTO in the $z_{850}$-band is taken as the total magnitude. The limiting magnitude for each band is estimated from the distribution of TFIT photometric errors ($1\sigma$) for all galaxies in the catalog, which typically exhibits a well-defined peak. The peak value of the error distribution represents the typical uncertainty given the sensitivity of the data. Whenever TFIT fluxes are consistent with zero within errors, we take the 1$\sigma$ errors as upper flux limits. 

Finally, we have compared the GOODS TFIT catalog with the public GOODS-MUSIC catalog (v2.0) discussed in \citet{santini09}, which uses a similar algorithm (CONPHOT) to deal with source blending \citep{desantis07}. The photometry from the two catalogs are found to be consistent with each other with the median offset less than 0.1 mag down to 26.0 mag (for IRAC [3.6$\mu$m]). The scatter, however, is quite large ($\approx 0.27$ mag) towards the faintest bin. While the origin of the relatively large scatter is beyond the scope of this work as it is difficult to quantify without detailed analyses coupled with simulations, no systematic bias found between the two catalogs implies that our main results will remain robust against the particular choice of algorithms adopted for optical photometry discussed above. 

\section{Measuring Stellar Masses of High-Redshift Galaxies}

\subsection{SED Fitting}\label{sed_fitting}
Stellar population synthesis models are used to derive stellar masses of galaxies in our sample.  We use the updated version of the stellar population synthesis models of \citet[][ known as the BC03 model]{bc03}, which we refer to as the CB07 code hereafter. We use the CB07 models to compute the broadband color evolution of galaxies with different star formation histories and ages. We use a \citet{chabrier03} initial mass function (IMF) with a lower and upper mass cutoff at 0.1$M_\odot$ and 100$M_\odot$, respectively, while the metallicity is fixed to solar. While the CB07 model includes enhanced contribution from the TP-AGB stars \citep[see, e.g.,][]{maraston05}, it returns stellar masses very similar to the BC03 because the light from UV-selected galaxies is dominated by younger OB stars even at the rest-frame optical wavelengths sampled by the IRAC channels at $z>3$. The TP-AGB stars would contribute more at longer wavelengths (rest-frame near-infrared) and for older stellar populations \citep[$>1$ Gyr, also see Figure 5 in][]{stark09}. 

The model galaxy SEDs are redshifted in the range of $z=0.2-8$ with $\Delta z=0.1$. Internal dust extinction is computed according to the prescription of \citet{calzetti00} law in the range of $E(B-V)=0.0-0.95$ with $\Delta E(B-V)=0.025$. We also attenuated the resulting SEDs from the neutral hydrogen absorption in the intergalactic medium using the \citet{madau95} prescription.  The age of our model SEDs extends from 5 Myr to the age of the universe at the given redshift. The star formation history is parametrized as an exponentially decreasing star formation rate (SFR), where $\tau$ represents the $e$-folding decay time. We use $\tau=0.1,0.2,0.3,0.4,0.6,0.8,$ and $1.0$ Gyr, and constant star formation (CSF; $\tau\rightarrow\infty$). We precompute a large number of model SEDs spanning the parameter space described above, then integrate them through the filter response functions of all the observed bands ($HST$/ACS: $B_{435}V_{606}i_{775}z_{850}$, {\it Spitzer}/IRAC: [3.6$\mu$m], [4.5$\mu$m], [5.8$\mu$m], VLT/ISAAC:$JHK_S$, CFHT/WIRCAM: $JK_S$, KPNO/MOSAIC $U$-band, and VIMOS/$U$-band) to compute the galaxy colors for each point in our model grid.

Using the precomputed model grid, we find the best-fit parameters for each galaxy in our sample. To evaluate the goodness-of-the-fit, we first determine the overall normalization to match the observed photometry, then compute the chi-square using all relevant data points.  Because model SEDs are normalized to have total mass of $1M_\odot$, the normalization factor directly determines the stellar mass. For each galaxy, we also determine the absolute UV magnitude at 1700 \AA, $M_{1700}$, by interpolating the best-fit model spectra. The evaluation of the chi-square value is done in flux units rather than magnitudes, which helps properly handling the data points with non-detection. Whenever the photometry in a given  passband is consistent with zero within the $1\sigma$ error, we assign as a flux upper limit a value that is representative of the sensitivity limit for the band. This treatment is often useful to avoid a large stellar mass to be assigned as a solution when TFIT returns very large photometric uncertainties. Such cases can occur when a source is hopelessly blended with much brighter neighbors. However, in most cases, using TFIT-returned flux and flux errors (instead of flux upper limits) for the fitting should yield similar results.  Data points with upper limits are included in the chi-square evaluation only when the model flux is higher than the upper limit. 

The flux upper limits in these bands are listed in Table \ref{tbl_1}, which are similar but not identical to those given in \citet{stark09} for the same data. The main reasons for this slight difference are two-fold. First, we used the source-weighted RMS map rather than the exposure map, which helps make more realistic estimate of the noise near bright sources. Second, TFIT includes the flux uncertainties arising from source blending, which, in particular, for faint sources like most high-redshift galaxies, is an important factor to account for as the majority of them are blended with brighter neighbors.

Instead of using the conventional approach of finding a single value as the best-fit stellar mass, we determine the full probability distribution function for each galaxy. In practice, this means that we retain all models that are within $\Delta\chi^2_{r}=1$ from the minimum $\chi^2$ model  as acceptable fits, and thus the stellar mass of the galaxy is described by a histogram of stellar masses spanned by all acceptable models. While this approach return results that are fully consistent with the conventional method when stellar mass of a given galaxy is in question, it has several advantages from a statistical standpoint. For example, there can be multiple models with $\chi^2$ values very close to one another but over a range of stellar mass. These situations become very real when one is dealing with galaxies at the low-mass end. Picking one model over the others may bias the outcome in an unpredictable fashion unless the systematics that affect the fitting are well understood. Choosing to construct the number counts using the full probability distribution makes it easier to correct the ``observed'' number counts for the measurement biases and systematics, which can be quantified using extensive simulations of galaxies. In \S4, we present our simulation procedures in detail, and further discussions on how best to measure the galaxy counts in stellar mass bins given the photometric uncertainties. 

The redshift distributions and the accuracy of photometric redshifts for our samples are well understood \citep{vanzella09, dahlen10}. Thus, we rely on photometric redshift estimation to derive physical parameters of each galaxy. During the SED fitting, redshift is varied within the best-fit photometric redshift by $\pm\Delta z=0.125$, which accounts for a typical uncertainty of the photometric redshift estimation \citep[see Figure 9 in][]{dahlen10}, except when the accurate spectroscopic redshift is known. Varying the redshifts is important in obtaining realistic errors in the derived SED parameters for these galaxies as the majority (95\% of the full $B_{435}$-band dropout sample) does not have a secure spectroscopic redshift. Typically, the errors in stellar mass when the redshift is varied are 20-25\% larger than those when the redshift is fixed to the best-fit photometric redshift. 

\begin{figure*}[t]
\epsscale{1.15}
\plottwo{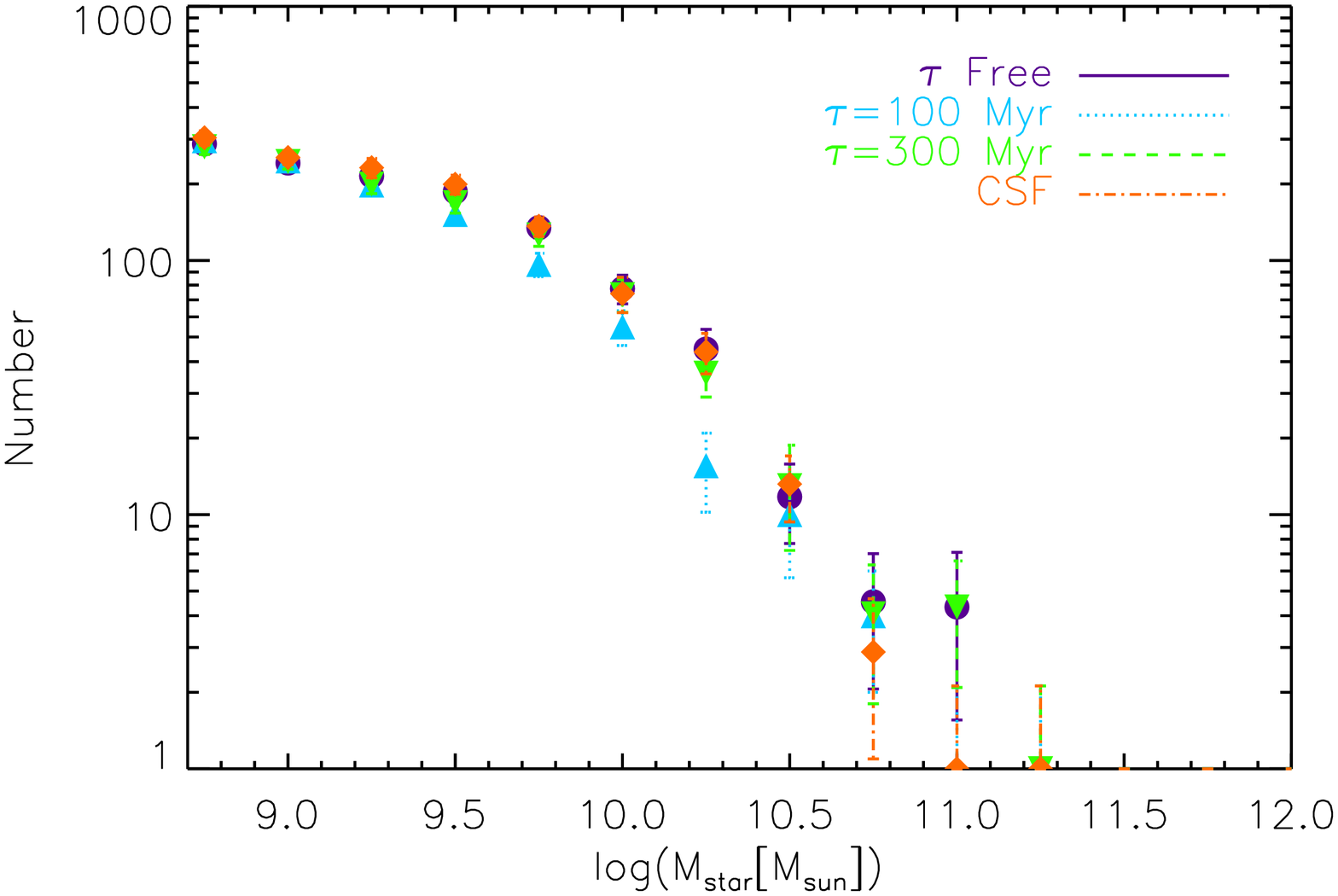}{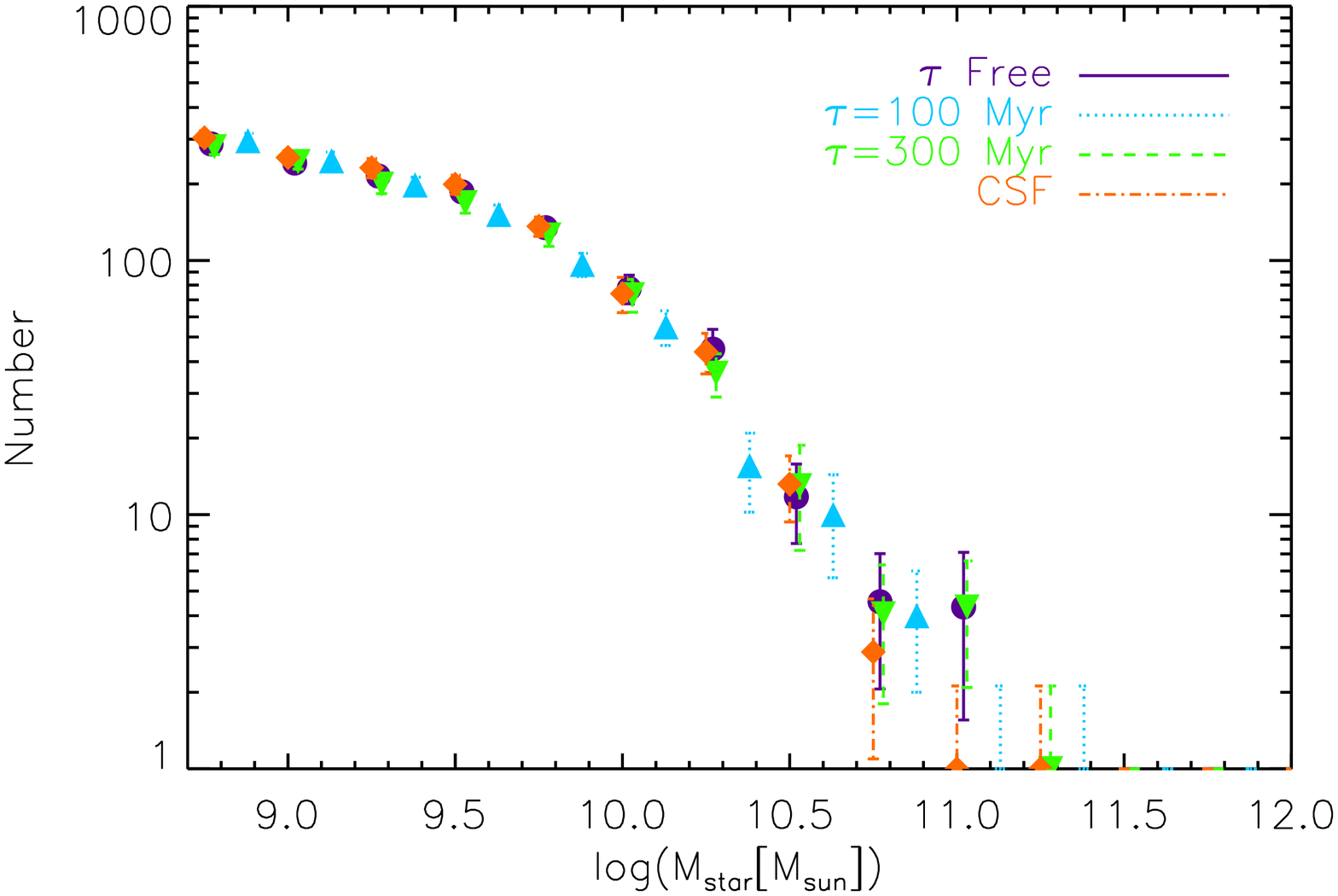}
\caption[2d_nc]{
Left: The observed number counts of LBGs at $z\sim4$ in stellar mass bins when different star formation histories are assumed. Larger $\tau$ values generally return higher stellar-mass-to-UV-light-ratios, and thus slightly larger masses for given rest-frame optical fluxes. Right: When a small offset is applied to the masses to correct for this effect, the number counts using different SFHs agree with one another very well. This means that even if mass measurements are biased towards lower or higher-masses because of our assumption in star formation histories, the overall shape of the number counts is likely robust. 
}
\label{sfhs}
\end{figure*}

\subsection{The Observed Number Counts of Galaxies}\label{n_obs}
Following the procedure described above, we measured the stellar masses of the $B_{435}$- and $V_{606}$-band dropouts and constructed the number counts. In order to assess robustness of our results, we investigate how the final number counts depend on the assumed star formation history. In Figure \ref{sfhs} we compare the number counts in the stellar mass bins for the $B_{435}$-band dropouts when different star formation histories are assumed. While we only include constant star formation history (CSF) and exponentially declining $\tau$ models in our fitting procedure, comparing the results using a short $\tau$ model (100 Myr, in this case) and CSF model gives an idea of how the results would differ for scenarios in which star formation is assumed to take place on short, "bursty" timescales, versus longer, more continuous modes.

We find that the observed number of galaxies in a given mass bin is slightly higher when a larger $\tau$ value is assumed as shown in the left panel of Figure \ref{sfhs}. This is because models with a larger $\tau$ typically result in a higher stellar-mass-to-UV-light ratio (mass-to-light ratio, hereafter), and thus predict slightly larger stellar mass for given rest-frame optical fluxes. However, different mass-to-light ratios seem to result in a solid shift rather than change in the overall shape of the number counts. They agree well with one another when a small offset is applied to correct for different mass-to-light ratios. In the right panel of Figure \ref{sfhs}, we show the same number counts with a small offset $\Delta \log [M_{\rm{star}}/M_\odot]=0,0.13,0.03,0.0$ for the free-$\tau$ model, $\tau$=100, 300, and CSF model, respectively. This shows that the overall shape of the number counts can be measured robustly regardless the assumed star formation history. One possibility is that  a typical $\tau$ value depends systematically on galaxy UV luminosity or stellar mass. For example, if low-mass galaxies have, on average, shorter SF timescales, such a dependency would manifest as a slightly steeper slope on the low-mass end. While it is difficult to model such an effect without more observational or theoretical input, Figure \ref{sfhs} suggests that such an effect should have a relatively small impact on the galaxy counts. Throughout this work, we use the number counts measured with ``free-$\tau$'' models for our analyses unless stated otherwise. 

There are several factors that may contribute, to a varying degree, to the estimation of the number counts. First, a hidden population of old stars formed from a previous generation of star formation may contribute to the stellar mass. The upper limit on this contribution can be set by assuming a population formed from an instantaneous burst at a very high redshift, which maximizes the mass-to-light ratio. Based on the multi-wavelength data set (out to $K$-band, which corresponds to the rest-frame $V$-band) for $z\sim3$ galaxies selected from the HDF-N,  \citet{papovich01} estimated that galaxies, on average, can be hiding stellar mass up to a factor of 3. While this can certainly be the case for some galaxies in our sample, the relative contribution from an old population is likely smaller because the universe is younger at $z\sim4$ and $\sim5$, and thus, the galaxies have had less time to have formed a significant stellar population unrelated to the current star formation. Furthermore, the {\it Spitzer} data samples out to, at least, the rest-frame $z$ ($I$)-band, $\approx$9000 (7500) \AA, at $z=4$ (5) as the majority of the galaxies with stellar mass $M_{\rm{star}}\geq 10^{8.7}M_\odot$ are detected in the [$3.6\mu$m] and [$4.5\mu$m] bands. For galaxies which are not formally detected in the {\it Spitzer} bands, it is difficult to assess how much mass may be hidden in old stellar populations. However, the average SED of {\it Spitzer}-faint galaxies constructed by  stacking \citep[see Figure 8 of][]{stark09} suggests that it is unlikely that such a contribution is more substantial than their more luminous counterparts. Furthermore, a large population of maximally old galaxies would be inconsistent with the rapid redshift evolution of the UV star formation rate density and UV luminosity function observed at $z\gtrsim4$ \citep{bouwens07, bouwens08}. 

Another interesting possibility concerns a recent result by \citet{shim11}, who reported a strong H$\alpha$ line emission in a large fraction ($>$70\%) of galaxies detected in the {\it Spitzer} [3.6$\mu$m] and [4.5$\mu$m] with high signal-to-noise ratios. While the study is based on a relatively small sample (70 galaxies) of spectroscopically confirmed sources at $3.8<z<5.0$, they convincingly showed that the flux excess in the {\it Spitzer} [3.6$\mu$m] band is indeed due to strong H$\alpha$ emission from star formation and not from hidden AGN activities. From the viewpoint of measuring stellar mass using broadband photometry, the presence of strong emission line can skew the mass estimation because emission lines are not included in our SED modeling. Furthermore, the observed equivalent widths of the H$\alpha$ emission, they claimed, suggest that it requires extremely young ages, top-heavy IMF, or sub-solar metallicity (or a combination of these). In practice, IRAC [3.6$\mu$m] is one of the main drivers in the SED fitting procedure that determines the stellar mass as it samples the rest-frame $I$-band. We visually inspected the observed SEDs of the $B_{435}$-band dropout sample to check whether such an excess is common also in the photometric sample. While the effect of line emission in the photometric data is fairly subtle to identify visually, and may be particularly hard to see for the fainter galaxies without spectroscopic redshifts,
we determine that the number of sources with a noticeably large [3.6$\mu$m] excess is indeed small, and thus unlikely result in change in the overall shape of the galaxy number counts. Because the strength of H$\alpha$ emission line should be lower for UV-faint (lower SFR) galaxies (a typical $z_{850}$-band magnitude for \citet{shim11} sample is 25.0, while the majority ($>95$\%) of the galaxies in our sample are fainter than 25.0 mag), the lack of H$\alpha$-excess sources seems to be in line with a possibility that the line emission is indeed due to star formation. Regardless of the nature of the flux excess reported by \citet{shim11}, we conclude  that it is unlikely that it will change our results significantly.

\subsection{The Observed $M_{UV}$-$M_{\rm{star}}$ Scaling Law}
\begin{figure*}[t]
\epsscale{1.15}
\plottwo{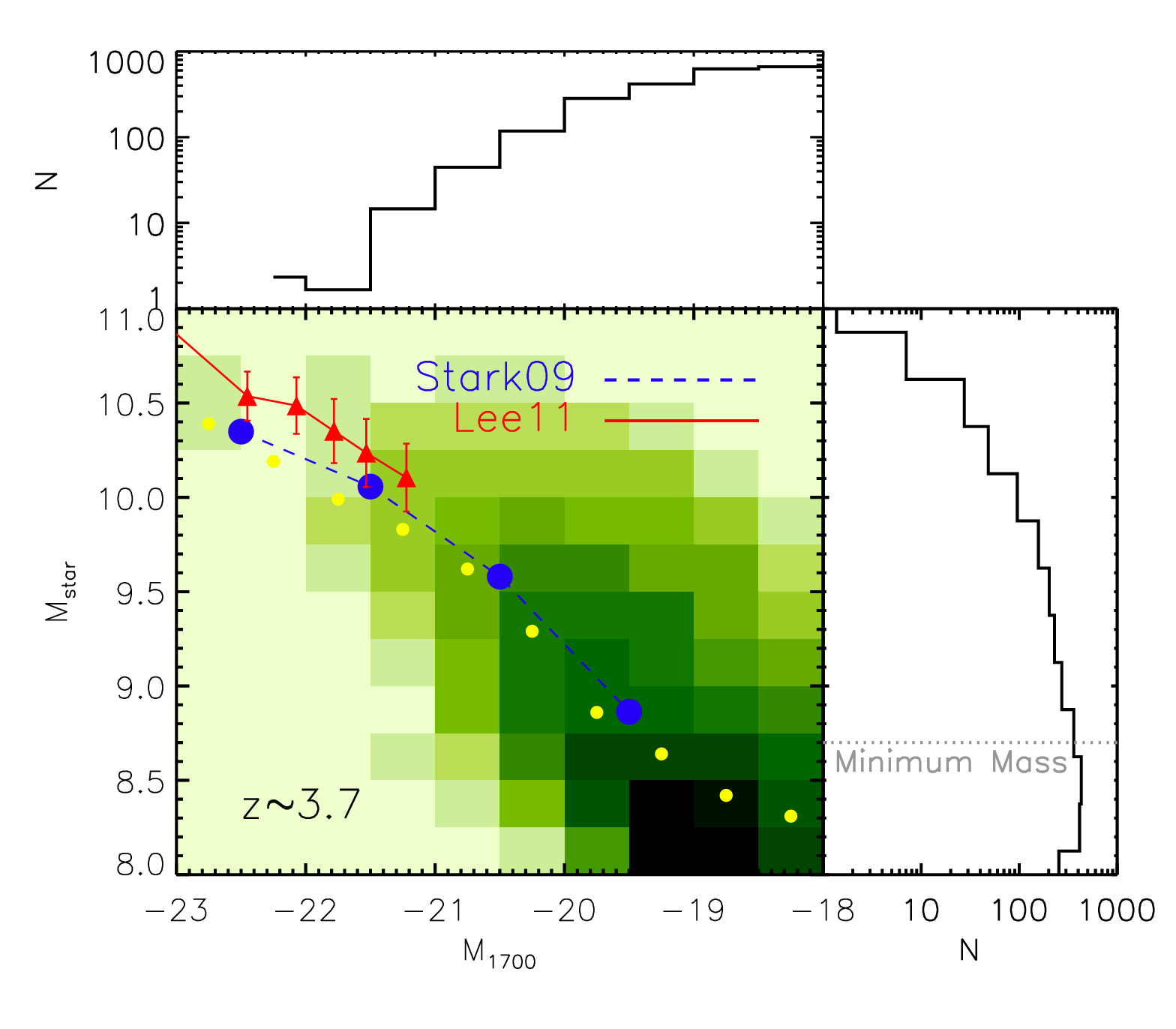}{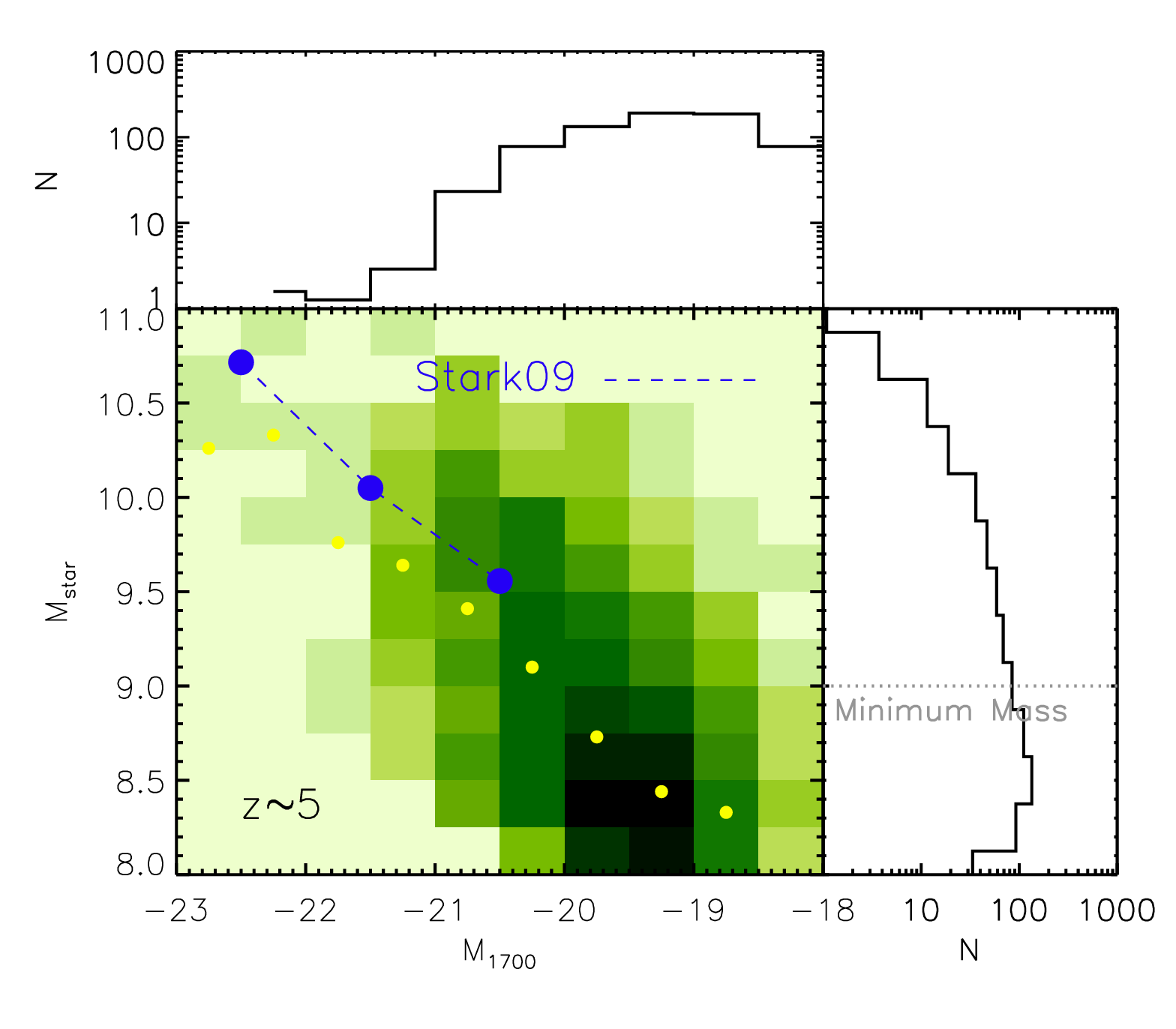}
\caption[2d_nc]{
The number counts of LBGs in UV luminosity and stellar mass bins at $z\sim4$ (left) and $z\sim5$ (right). We show in the main panels, the location of all galaxies in our sample. Each cell is color-coded with the observed number of galaxies. The average stellar mass at a fixed UV magnitude is also shown as yellow filled symbols. Our results  indicate an overall correlation between UV luminosity and stellar mass albeit with a significant scatter.  Our estimates are in good agreement with those of \citet[][]{stark09} as marked by dashed line and filled symbols, but extend further down to lower masses using TFIT photometry. For $z\sim4$, we also show similar measurements by \citet{lee11} for very UV-luminous ($M_{1700}<-21$) galaxies. In upper panels and right panels, we show the galaxy number counts in UV magnitudes and stellar mass bins, respectively. The minimum mass we include in our stellar mass function analyses, which we determine based on our simulations (\S \ref{sim}), is also indicated on right panels. Our number counts in UV magnitudes are in excellent agreement with \citet{bouwens07}, confirming that the UV luminosity function has a steep faint-end slope. A large concentration of UV-faint galaxies in low masses is due largely to non-detection in IRAC channels and thus represents upper mass limits in some galaxies. This suggests that at $M_{1700}>-19$ the median stellar mass for galaxies falls more steeply than that for more luminous ones. 
}
\label{2d_nc}
\end{figure*}

Having measured both UV luminosity and stellar mass of the dropout samples, we present how these galaxies populate  the plane of the observed luminosity and stellar mass in the main panels of Figure \ref{2d_nc}. The figure is color-coded by the total number of galaxies in each given cell, and thus illustrates the overall trend of the relative location of galaxies on the plane as well as the number of galaxies at different mass and luminosity bins. The average stellar mass at a fixed UV luminosity is also computed from our measurements as shown as yellow filled circles. Top and right panels indicate the sum of all columns and rows (of the main panels), i.e., galaxy counts as a function of UV luminosity and stellar mass, respectively. We also note that a large concentration of UV-faint galaxies at very low masses ($10^{8.0}-10^{8.5}M_\odot$) is partly an artifact produced by a combination of flux upper limits and our adopted fitting method. When a source is not formally detected in the {\it Spitzer} IRAC channels, the minimum $\chi^2$ is achieved when the model flux is below all the flux upper limits set by the data. As a result, the best-fit solution is returned with a range of low stellar masses, which is primarily determined by a range of stellar-mass-to-UV-light ratios allowed within our model grid (including the models with the lowest mass-to-light ratios as the data does not provide any constraints), and thus does not reflect a true value. To avoid any systematic bias in our results by this effect, we confine our analyses only to galaxies with stellar mass  $M_{\rm{star}}\geq10^{8.7}M_\odot$ for the $B_{435}$-band dropouts and $M_{\rm{star}}\geq10^{9}M_\odot$ for the $V_{606}$-band dropouts. These limits, the minimum stellar mass considered in our analyses, are also indicated in Figure \ref{2d_nc} on right panels as ``Minimum Mass''. Our simulations (\S\ref{sim}) indicate that the current data allow statistically robust measurements down to this limit.  A majority of simulated galaxies in this mass range are formally undetected in the IRAC bands, or have large photometric errors. 

Our measures are largely consistent with those made by \citet{stark09} marked by dashed lines, but are generally slightly lower (by $\sim0.1$ dex) than their estimate. The discrepancy may be due to the fact that they adopted a single value for each stellar mass rather than taking into account the full distribution, and as a result, the median is slightly skewed towards higher masses. As we will show in \S\ref{sim}, the uncertainty in the mass determination at $10^{9}M_\odot$ is somewhat large, and thus requires statistical approach to construct the distribution. Furthermore, the discrepancy may reflect the difference in the two samples,  that ours includes a much larger number of galaxies ($\approx$90\% of the photometric sample in comparison to their 35\%) because TFIT photometry does not require that sources be well isolated. 


In Figure \ref{2d_nc}, we show that  there is an overall correlation between stellar mass and UV luminosity down to $M_{\rm{star}}\approx10^{8.7}M_\odot$ in that more UV-luminous galaxies are on average more massive. However, it is equally evident that there is a substantial scatter to this correlation. In particular, there exist galaxies that are quite massive but faint in the UV.  Roughly 36\%  and 20\% of the galaxies in the $B_{435}$- and $V_{606}$-band dropout sample that are more massive than $M_{\rm{star}}=10^{9.5}M_\odot$ are fainter than $M_{1700}=-20$. The implication is that there is a non-negligible population of galaxies whose SFRs were considerably higher in the past, but are currently fading in the UV. It is possible that some of them may be faint simply due to more dust reddening, but as we will show later (\S\ref{M1700_Mstar}), dust alone is not enough to explain the observed UV luminosity distribution of these low-mass galaxies.  Interestingly, the opposite is not true in that the region of the low stellar mass and high UV luminosity is largely devoid of galaxies. This suggests that the majority of the UV-luminous galaxies have been forming stars for quite some time (at least a few hundred million years). This interpretation is consistent with a recent study of the most UV-luminous galaxies at the same redshift by \citet[][also shown in Figure \ref{2d_nc}]{lee11} who concluded that they undergo smooth star formation histories \citep[also see,][]{shim11}.

\section{Understanding Measurement Uncertainties/Biases}\label{sim}
Spectral energy distribution (SED) fitting is a procedure that attempts to fit a number of galaxy parameters based on multiple data points taken at different depths and resolutions. Thus, its outcome may depend strongly on the range of  galaxy parameters (initial mass function, star formation histories, age) and the quality of the photometric data. Therefore it is worth paying close attention to how well each galaxy parameter can be measured given the set of observations. The most relavant issues to our analyses are: How robust are the derived galaxy properties (e.g., extinction, stellar mass) when the SFH and redshift are unknown? What are their uncertainties and how do they scale with the quality of photometry? How well can we measure the galaxy light within the same effective aperture from the multi-resolution data set, using a photometric technique of choice (e.g., SExtractor, TFIT)? Creating a mock data set with realistic properties (noise, seeing, galaxy colors, etc.) is key to properly accounting for all factors that may contribute to the measurement errors and systematics. To address these questions, we carried out a large set of simulations. The procedure is designed to simulate galaxies as realistically as possible in every step of the way by mimicking the  noise properties, size distribution, photometric extraction, and derivation of physical parameters via SED fitting, of the real high-redshift galaxies.   

\subsection{Galaxy Simulations on Mixed Resolution Dataset}
The simulation procedure can be summarized as follows: 1) Create galaxy model SEDs spanning a wide range of physical parameters (age, reddening, SFHs, redshift, and stellar mass), 2) Compute input photometry of these SEDs in the observed passbands, 3)  Insert mock galaxies into the images in the different photometric bands, 4) Measure photometry using TFIT software, and 5) Carry out SED fitting to ``measure'' output physical parameters. By design, the output of our simulations is identical to that of the real galaxies except that input parameters are known for simulations. By comparing the input and output parameters, we can quantitatively assess the robustness of the measurements at various levels and use that information to correct for the systematics. As for the range of physical parameters used to create mock galaxies, we use the same ranges that we explore in the SED fitting procedure for real galaxies. By doing so, we ensure that we can quantify all the measurement errors and biases within the assumptions that went into deriving the physical parameters of real galaxies. Any non-standard physical characteristics of real galaxies, such as exotic IMFs or sub- or super-solar metallicities, that may have contributed to our measurements, however, cannot be accounted for in our simulations, and thus is beyond the scope of this work. The descriptions of the steps \#1 and \#2 are given in \S\ref{sed_fitting}. The details of inserting mock galaxies in the real data (\#3) are found \citet{ferguson04}. 

The main improvement  of our simulations over the previous version described in \citet{ferguson04} is that we extend the technique to create multi-resolution data sets. This revision is straightforward as all of the data have pixel scales that are integer-multiples of one another (0\farcs03/pix for the $HST$/ACS data, 0\farcs5/pix or 0\farcs30/pix for the ground-based data, and 0\farcs60/pix for the {\it Spitzer}/IRAC data). All images are initially created in the highest resolution, convolved with the appropriate PSFs, then block-summed to the desired pixel scale before being added to the real science images (step \#3). Only 30 galaxies are inserted at a time over a 4 arcmin$^2$ area (4096 $\times$ 4096 in the pixel resolution of the ACS bands) to avoid unrealistic crowding. They are placed at random locations within the images with random orientation angle. The light profiles of galaxies are chosen to be either of De Vaucouleurs profile or of exponential disk. The size distribution is set to follow a lognormal distribution as observed at these redshifts \citep{ferguson04}.  

For each run, once the mock images are created, we repeat the identical procedures for measuring their photometry (step \#4) and SED parameters (step \#5) as for real galaxies. This means that we run SExtractor on the ACS images with the same setup for source detection and photometry, then use the ACS $z_{850}$-band as templates to run TFIT on other passbands. The $U$-band (below the Lyman limit for $z>3$ galaxies) was not included in the simulations, as mock galaxies do not suffer from low-redshift interlopers, for which the $U$-band is mainly used to check non-detection for real galaxies. 

\subsection{How Well Can We Measure Galaxy Parameters?}\label{galsim}
\begin{figure*}[t]
\epsscale{0.9}
\plotone{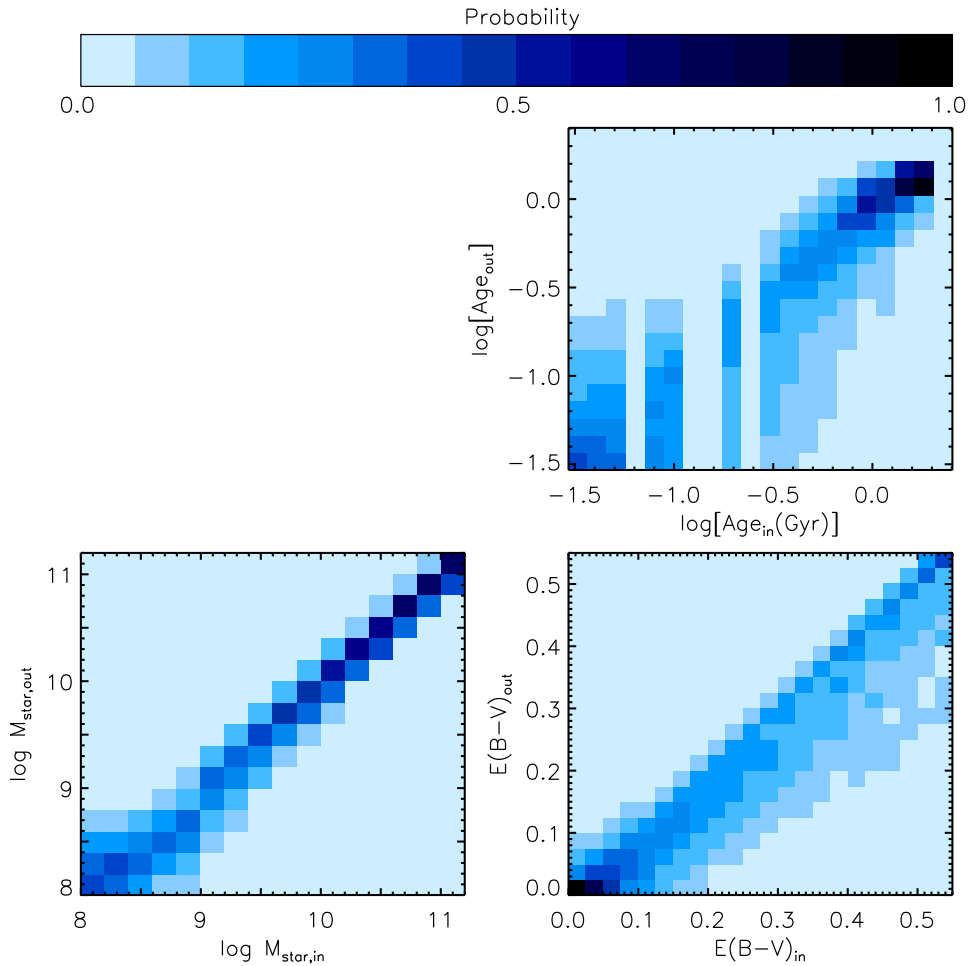}
\caption[plot_sedfit_in_out]{
The probability distribution function for the stellar mass (lower left), internal reddening (lower right), and age parameters (top right) recovered from our SED fitting procedure as a function of input parameters.  The color bar (top) indicates the probability value for each (input, output) cell. The discreteness of input ages reflects non-equal steps of ages adopted for the simulated galaxies. These results are based on 34,501 $B_{435}$-band dropouts planted in the GOODS-S and GOODS-N field according to the simulations procedure we described in text (\S\ref{sim}). Based on these results, we determined that stellar mass is the most robust quantity that can be recovered, at least, based on the broad-band photometry via SED fitting. The internal reddening can be recovered to a somewhat lesser degree, indicated by a larger error. The population age is the least robust parameter, which is likely driven by the current sensitivity limit of the near-infrared data sampling the Balmer/4000\AA\ region ($H$ and $K_S$ band for the $B_{435}$-band dropouts). 
}
\label{plot_sedfit_in_out}
\end{figure*}

Using the procedures described above, we have simulated about 200,000 galaxies, of which $\approx$35,000 galaxies are selected as $B_{435}$-band dropouts and $\approx$26,000 as $V_{606}$-band dropouts using the color selection criteria  given in Equation 1 and 2.  These galaxies span a wide range of UV luminosities, star formation histories, age, extinction, and stellar mass. In Figure \ref{plot_sedfit_in_out}, we show the input and recovered value of population age, dust reddening (parametrized as $E(B-V)$), and stellar mass, for simulated galaxies. Stellar mass is the most robust quantity that we can recover based on the available broadband photometry at the current depth, while internal dust reddening can be recovered to a somewhat lesser degree (bottom right). As for population ages, our SED fitting fares very poorly with a large scatter (top right) unless the age of input galaxies is much older ($>500$ Myr) than the values derived by SED fitting for the majority of the observed Lyman break galaxies. The poor recovery of population age is likely attributed to the current (shallow) depth of the near-infrared data from ground, which samples near the Balmer/4000\AA\ break ($H$ and $K_S$ band for the $B_{435}$-band dropouts at $z\sim4$). These results are in qualitative agreement with \citet{joshualee09}, who, based on the SED fitting of mock galaxies created by a semianalytic model, reported  that stellar mass can be recovered most reliably (within 0.1 dex of the true values) even when the assumed star formation history is clearly different from the intrinsic one. 

\begin{figure*}[t]
\epsscale{0.7}
\plotone{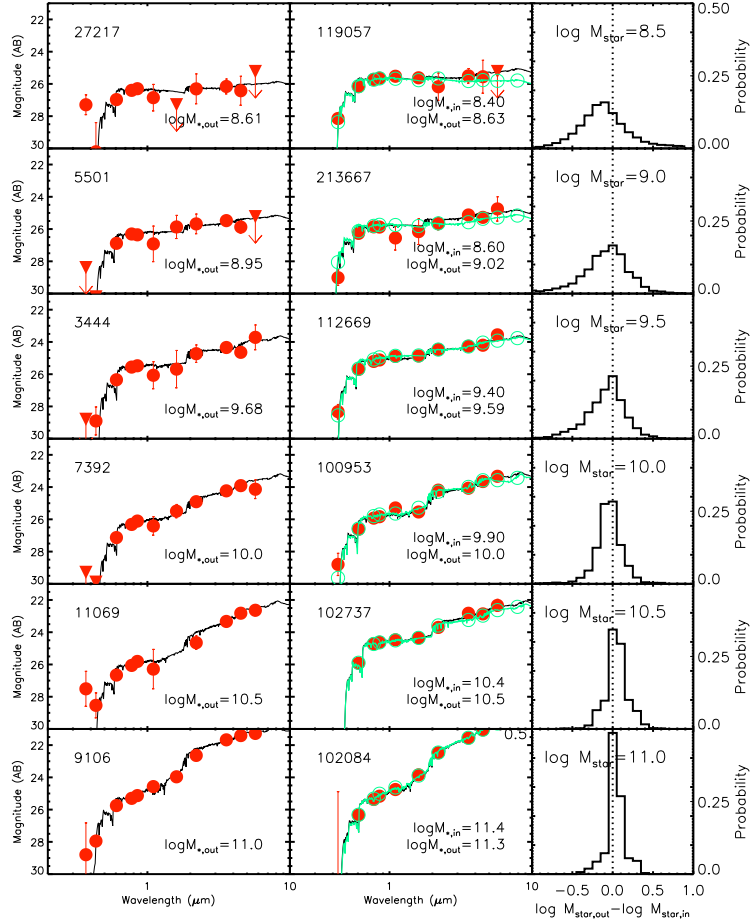}
\caption[plot_seds]{
We show the spectral energy distributions of typical galaxies at different stellar mass bins for both {\it real} (left panels) and {\it simulated} galaxies (middle panels). Filled circles represent flux in each photometric band with TFIT errors, while solid lines indicate the best-fit population models from which the stellar mass is determined. All stellar masses are given in units of solar mass. In the histograms on right panel we show the distribution of {\it recovered stellar mass} with respect to the {\it input} mass. The typical accuracy in the determination of stellar mass can be assessed based on the mean and dispersion of the output mass distribution. We conservatively set the mass limit to  $\log ~[M_{\rm{star}}/M_\odot] = 8.7$ and $9.0$ at $z\sim4$ and $5$, respectively. Below these limits, output masses are increasingly uncertain as evidenced by an offset with large scatter (top right panel). }
\label{plot_seds}
\end{figure*}

In Figure \ref{plot_seds}, we show typical SEDs of real (left) and simulated (right) galaxies at different stellar masses. For the simulated galaxies, we also indicate the input photometry as open circles, illustrating the increasing degree of photometric scatter and errors towards the low-mass end. On right panels, we show the probability distribution of the {\it recovered} stellar mass at various input mass values.  As expected, it is evident that the accuracy of stellar mass estimation decreases rapidly towards low masses as a result of larger photometric errors in relevant passbands (in this case, mainly driven by the IRAC [3.6$\mu$m] and [4.5$\mu$m]). Increasingly large spread of recovered mass with respect to the true value further supports the statistical approach we take in constructing galaxy counts in mass bins. In any stellar mass function that rises relatively steeply towards the low-mass end, such scatter will result in artificial steepening of the observed slope as more number of low-mass galaxies will scatter into higher-mass bins than the opposite. 

In our analyses throughout this paper, we use the mass probability distribution (right panels of Figure \ref{plot_seds}) for two purposes; first, to assess and thereby determine how much we can push the limit of the available data without introducing systematic biases into our measurements, second, to statistically correct the observed number count for the ``mass spread'' due to photometric errors. As for the former, we conclude that stellar mass can be measured reliably down to $\log [M_{\rm{star}}/M_\odot]=8.7$ and $9$ at $z\sim4$ and $5$, respectively. As evident in right panels of Figure \ref{plot_seds}, the probability distribution becomes skewed towards the masses below the limit. This is mainly due to the fact that sources become very  faint (generally, in all bands). This leads to two independent consequences that affect the SED fitting procedure. First, the SExtractor tends to underestimate the isophotal area from the detection image, which is later used to define the ``effective aperture'' in other bands to measure flux using TFIT. More quantitative discussions on this issue will be discussed in an upcoming paper (A. Galametz et al., in prep). Second, because the same sources are (usually) intrinsically fainter in the {\it Spitzer} bands, and as a result, the SED fitting procedure is often driven by the flux upper limits. Because there is no meaningful constraint on the rest-frame optical wavelengths, the fitting returns a wide distribution of stellar masses as acceptable (see discussions in \S\ref{n_obs}).

\section{The UV Luminosity Function and Stellar Mass Function of LBGs}\label{LF_SMF}
\subsection{The UV Luminosity Function}\label{UVLF}
We measure the UV luminosity function of LBGs in our samples using a methodology similar to that presented by \citet{bouwens07}. The procedure consists of 1) constructing the number counts of the real galaxies in apparent magnitude bins, 2) quantifying the photometric and selection completeness of the observed galaxies by using  simulations (\S\ref{sim}), and 3) determining the range of UVLF parameters that reproduces the observed number counts. In quantifying the completeness, we treat internal dust reddening as an explicit variable. 

We define the selection efficiency, $p_S(m',z,\epsilon)$,  as the probability that a galaxy at redshift $z$ of the (true) apparent magnitude $m'$ with the reddening parameter, $E(B-V)=\epsilon$, to be detected and selected as a dropout in our selection. In addition to the selection efficiency, we further quantify the photometric error that affects the number counts of the dropouts as the probability that a galaxy $(m',z,\epsilon)$ to be observed at the apparent magnitude $m$, hereafter referred to as $p_M(m|m',z,\epsilon)$. 
To relate the intrinsic UV luminosity function to the observed number density, we define the effective volume as:
\begin{eqnarray}
\label{veff}
V_{m,k} &=& \int_z \int_{m'} W(M(m',z,\epsilon)-M_k)\mathcal{P}(m|m',z,\epsilon)\frac{dV}{dz}dm'dz \nonumber \\
               &\approx& \sum_{m'} \sum_{z}W(M(m',z,\epsilon)-M_k)\mathcal{P}(m|m',z,\epsilon)\frac{dV}{dz}(z)  \nonumber\\
                && \times \Delta m' \Delta z
\end{eqnarray}
where $W$ is a window function, and $\mathcal{P}(m|m',z,\epsilon)\equiv p_S(m,z,\epsilon)\times p_M(m|m',z,\epsilon)$, $dV/dz$ is the cosmic volume covered by the survey area, $m$ is the {\it observed} apparent magnitude, $m'$ is the {\it true} apparent magnitude, and $M_k$ is the absolute UV magnitude at 1700 \AA. The effective volume averaged over the intrinsic distribution of reddening $\epsilon$ is:
\begin{equation}
\label{veff_reddening}
\langle V_{m,k} \rangle_\epsilon = \frac{\Sigma_n V_{m,k}\mathcal{D}(\epsilon_n,M_k)\Delta \epsilon}{\Sigma_n \mathcal{D}(\epsilon_n,M_k)\Delta \epsilon}
\end{equation}
where $\mathcal{D}$ characterizes the distribution of reddening which can be luminosity-dependent ($M_k$). The observed number count for a given luminosity function can be computed as:
\begin{equation}
\label{nc_m}
N_m= \sum_k \phi_{L}(M_k)\langle V_{m,k}\rangle_\epsilon
\end{equation}
where $\phi_L(M_k)$ is the UV luminosity function evaluated at the absolute magnitude bin $M_{1700}=M_k$, where the UVLF is parameterized as a Schechter function:
\begin{eqnarray}
\phi_L(M_k)&=&(0.4 \ln 10) \phi^* ~10^{0.4(M_{\rm{UV}}^*-M_k)(1+\alpha)}\nonumber \\
&&\times \exp[-10^{0.4(M_{\rm{UV}}^*-M_k)}) ]
\end{eqnarray}
As for the reddening distribution $\mathcal{D}$, we assumed a normal distribution with a mean of $\epsilon=0.14$ if $M_k\leq-21.06~ (L^*)$, $\epsilon=0$ if $M_k\geq-18.56 ~(0.1 L^*)$, and linearly declining in between. The $1\sigma$ scatter of the distribution remains $\sigma_\epsilon=0.14$ at all luminosities. Our prescription is  essentially identical to that used by \citet{bouwens07}. \\
\begin{figure*}[t]
\epsscale{1.0}
\plottwo{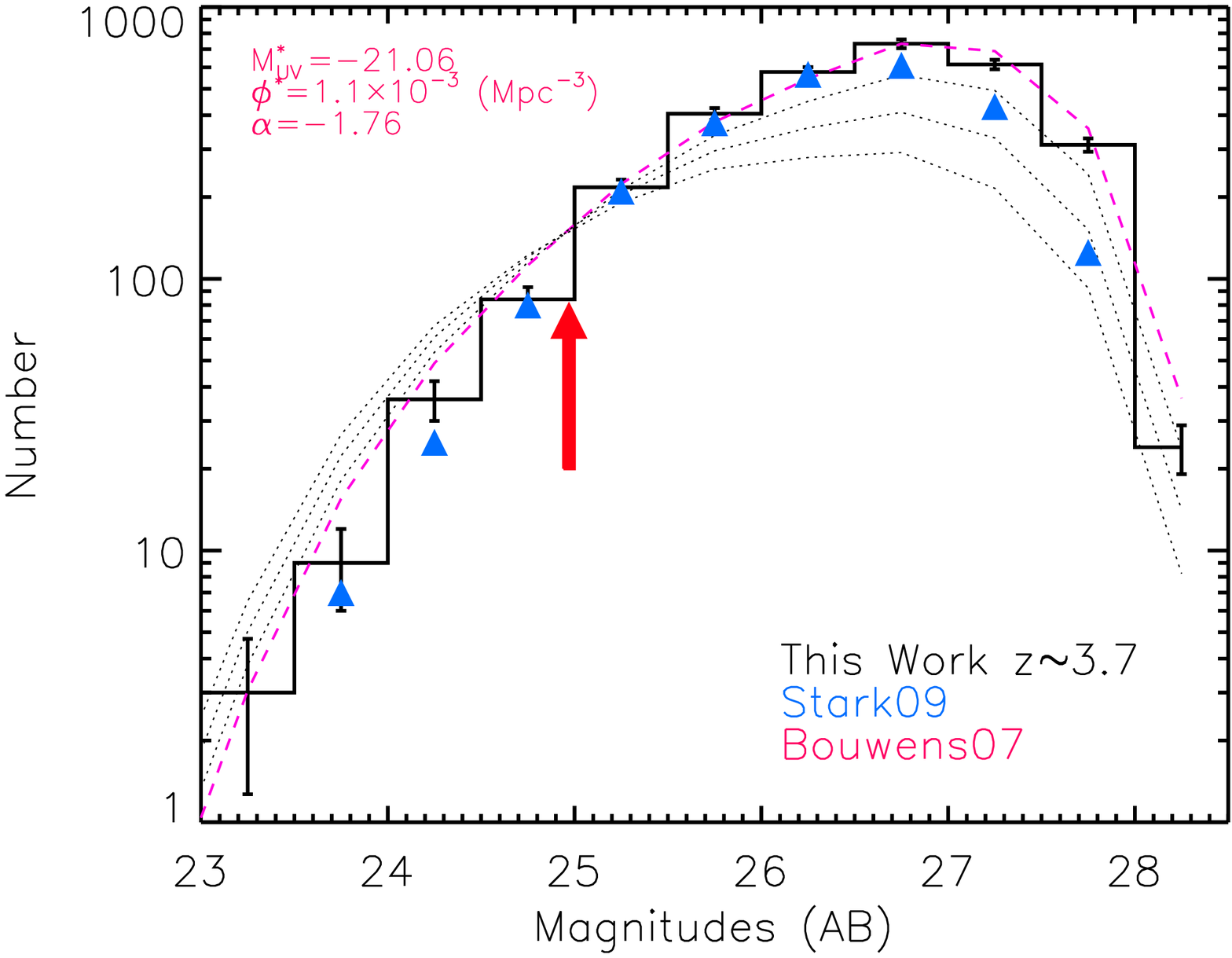}{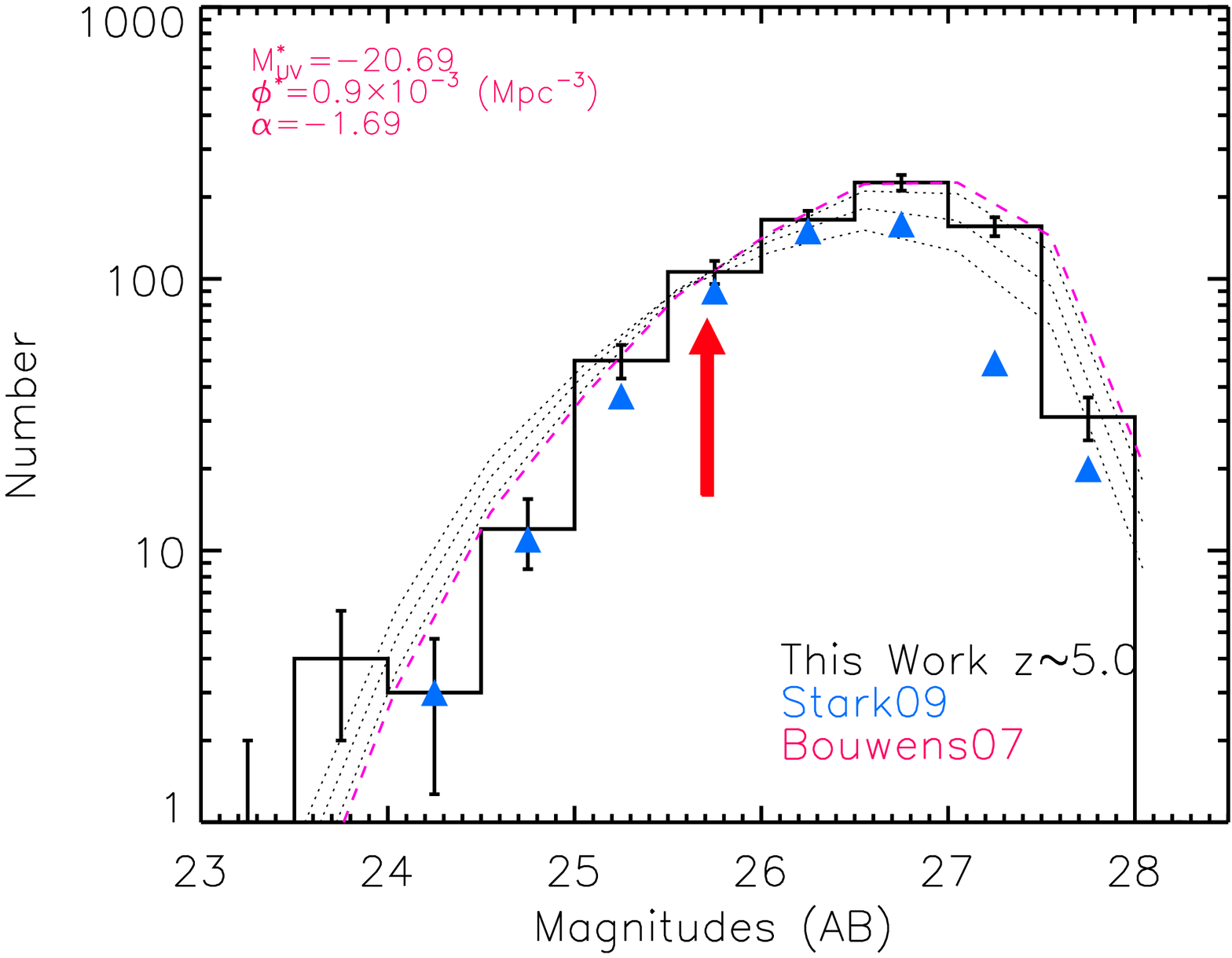}
\caption[plot_nc_mbins]{
The total {\it observed} number of galaxies in UV magnitude bins compared to those in the literature at $z\sim4$ (left) and $5$ (right). Our number counts are in excellent agreement with those by \citet{stark09} except the very faint end. This discrepancy is  likely explained by the fact that the data set (v2.0 release) used in this work has a slightly higher completeness than theirs (v1.0 release). Assuming the best-fit parameters of the \citet{bouwens07} determination of the UV luminosity function, we compute the expected galaxy number counts using Equation \ref{nc_m}. We find good agreement between the observed and expected number counts, which  confirms that the faint-end slope of the UV luminosity function is indeed steep. Also shown as dotted lines in each panel are the expected number counts when a set of shallower slopes ($\alpha=-1.6, -1.4, -1.2$ from top to bottom) are assumed while the characteristic luminosity $M_{1700}^*$ is fixed to the same value as shown in top left corner. All models are normalized at $z_{850}=24.97$ and $25.71$ at $z\sim4$ and $5$, roughly corresponding to the respective characteristic luminosity at the median redshift of each sample (shown as filled arrows). 
}
\label{plot_nc_mbins}
\end{figure*}

In Figure \ref{plot_nc_mbins}, we show the observed number counts together with those published in \citet{stark09} based on the GOODS v1.0 catalog. Both measurements agree well except in a few faintest magnitude bins, where our deeper catalog should be more complete. Thick dashed lines indicate the expected number counts computed according to Equation \ref{nc_m} when the best-fit UVLF parameters given by \citet{bouwens07} are assumed. While we are unable to independently determine the UVLF parameters based on the GOODS data alone, the good agreement between the two suggests that the faint-end slope is steep in the range of $-(1.7-1.8)$. To illustrate the fact that our data is indeed more consistent with a steep ($< -1.7$), we also show in Figure \ref{plot_nc_mbins} as a set of dotted lines the model number counts with the slope $-(1.6, 1.4, 1.2)$ when the characteristic luminosity $M_{\rm{UV}}^*$ is fixed. All functions are normalized at $z_{850}=24.97$ and $25.71$ at $z\sim4$ and $5$, roughly corresponding to the respective characteristic luminosity at the median redshift of each sample (shown as filled arrows).

In comparison with the results by \citet{bouwens07}, we find a slight discrepancy at the bright end, where their estimate is about a factor of 2 higher than what we and \citet{stark09} found. This is likely due to the fact that we discarded some number of bright sources after  inspecting their SEDs (as well as MIPS detection, see \S \ref{mips_sources}). Many of bright sources have UV colors consistent with high-redshift LBGs (and thus included in their sample), but longer wavelength data suggests that they are unlikely to be at high redshift. Clearly, both our estimates and \citet{bouwens07} suffer from small-number statistics at the bright end, and the discrepancy is driven by mere 10-20 sources and their photometric redshifts, more complete spectroscopy of these sources and larger-area surveys will be needed to better quantify the UVLF at the bright end. 

\subsection{The Stellar Mass Function of SFGs}\label{SMF}

Using robust measurements (and error estimates) of SED properties and UV luminosities for individual galaxies, we next determine the stellar mass function of LBGs at $z\sim4$ and 5. We caution readers that these measurements do not in any way represent the {\it total} stellar mass function of {\it all} galaxies, but rather provide  information on how much of the cosmic stellar mass budget is contributed by actively star-forming and relatively UV-bright galaxies. Furthermore, as we will see later, by comparing the shape of the SMF with that of the UV LF measured for the same galaxies, one can obtain crucial clues to how (UV-visible) star-formation has proceeded with cosmic time. 

We compute the number density of LBGs at each stellar mass bin as follows. First, the number density of galaxies at a fixed apparent magnitude $m_j$ and stellar mass $M_{*,m}$ is obtained by correcting the observed number $N_{{\rm obs}}(m_j,M_{*,m})$ for the selection incompleteness by dividing it by the effective volume $V_{\rm{eff}}(m_j)$.  Using Equation \ref{veff} and \ref{veff_reddening}, the total effective volume at a fixed apparent magnitude $m_j$ is 
\begin{equation}
\label{veff_smf}
V_{\rm{eff}}(m_j)=\Sigma_k \langle V_{j,k} \rangle_\epsilon
\end{equation}
\begin{equation}
\label{phim_jm}
\phi_M(m_j,M_{*,m})=\frac{N_{\rm{obs}}(m_j,M_{*,m})}{V_{\rm{eff}}(m_j)}
\end{equation}
The {\it total} number density of LBGs at a stellar mass bin is obtained by adding the contributions from all the apparent magnitude bins:
\begin{equation}
\phi_M(M_{*,m})=\Sigma_j \phi_M(m_j, M_{*,m})
\end{equation}
To estimate realistic errors at each mass bin, we account for both random Poisson errors (based on the observed number at each bin) as well as sample variance (based on the difference between the observed number at two fields). The Poisson error at each bin is computed as $\Delta \phi_M (m_j,M_{*,k}) = [\sqrt{{N_{\rm{obs}}(m_j,M_{*,k})}+0.75}+1]/{V_{\rm{eff}}(m_j)}$ using the approximation given by \citet{gehrels86}. The total Poisson error at each mass bin is the quadratic sum of all magnitude bins:
\begin{equation}
\Delta \phi^\mathcal{P}_M(M_{*,k})=\sqrt{\Sigma_j [\Delta \phi_M (m_j,M_{*,k})]^2}
\end{equation}
The errors associated with sample variance are estimated by accounting for the difference in the observed number between the two fields as:
\begin{equation}
\Delta \phi^\mathcal{SV}_M(M_{*,k})=|\phi^S_M(M_{*,m})-\phi^N_M(M_{*,m})|/2
\end{equation}
The total error budget is obtained as a quadratic sum of the random error and sample variance at each mass bin, and is given as $\Delta \phi_M=\sqrt{[\Delta \phi^\mathcal{P}_M]^2+[\Delta \phi^\mathcal{SV}_M]^2}$.

\begin{figure*}[t]
\epsscale{1.0}
\plottwo{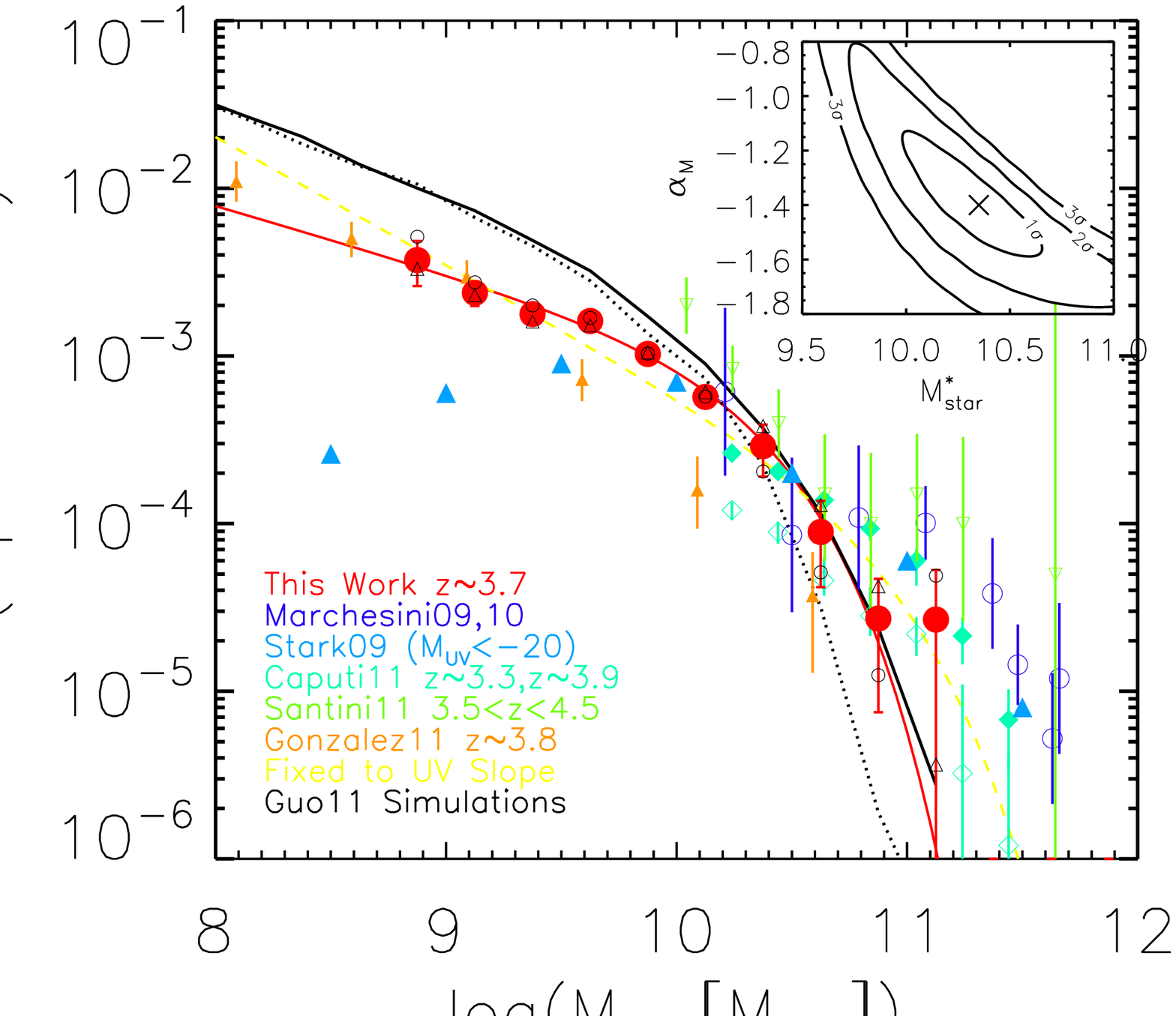}{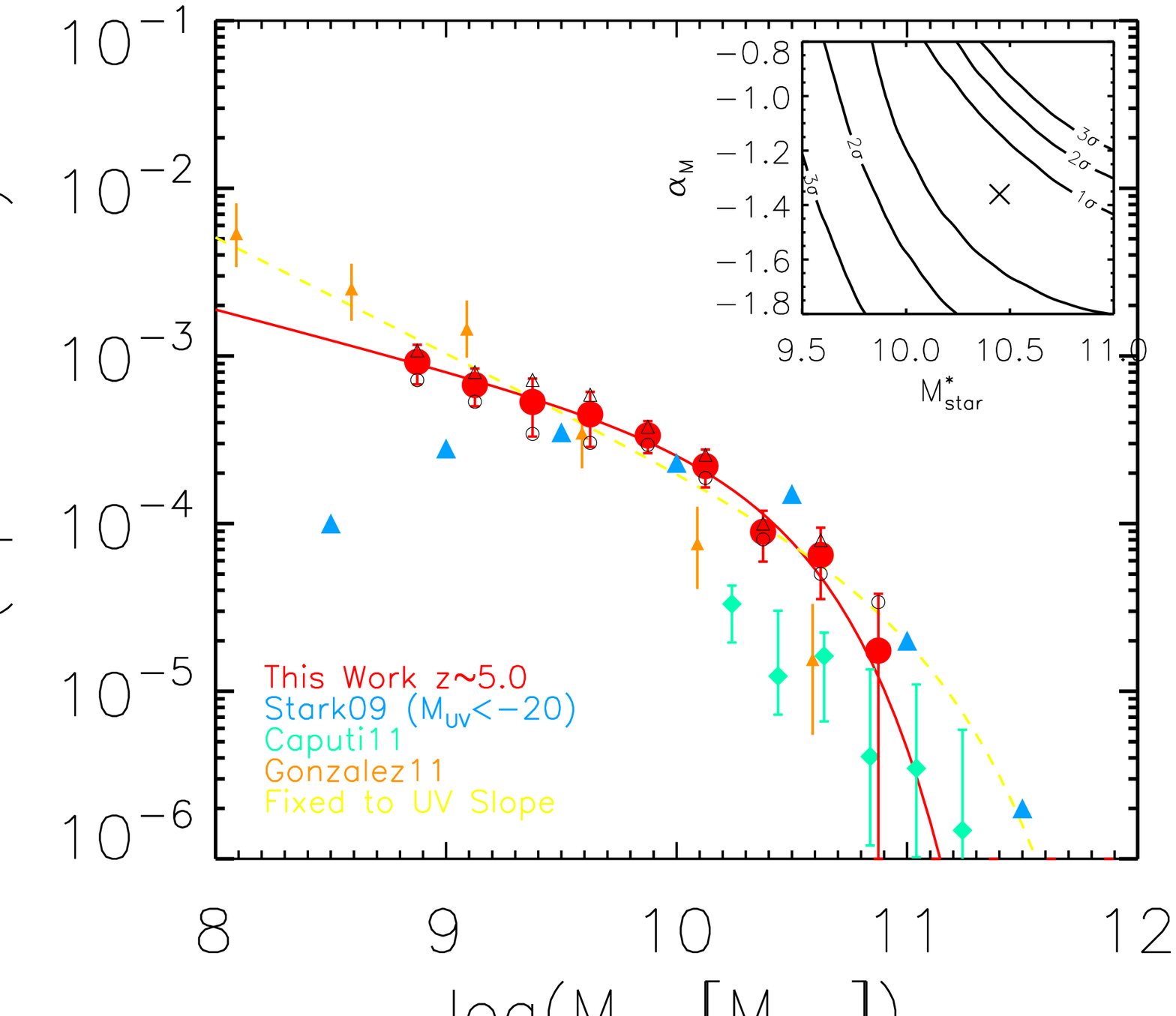}
\caption[plot_nc_lmbins]{
The stellar mass functions of LBGs at $z\sim4$ and $5$ are shown. The number density of galaxies at each stellar mass bin is obtained by first correcting the observed number for the selection incompleteness (for the UV selection), then summing up the contributions from all magnitude bins. The error bars fully account for random Poisson errors as well as those arising from sample variance. Filled circles represent our best estimates combining the statistics of both GOODS fields, while small open symbols (circle and triangle) at each mass bin indicate the number density computed in each field.  Also shown are similar measures in the literature \citep[][in blue filled triangles, blue open circles, cyan diamonds,  orange upward triangles, and green downward triangles, respectively]{stark09, marchesini09, marchesini10,caputi11,gonzalez11,santini12}. For $z\sim4$, we show the model predictions from \citet[][solid and dotted yellow lines]{guo11} based on Millennium and Millennium II simulations. The best-fit Schechter parameters (shown in solid red lines) indicate that the low-mass-end slope of the stellar mass function is $\approx -1.4$, and thus is considerably shallower than the faint-end slope of the UV luminosity function derived for the same galaxies, $\approx -(1.7-1.8)$. Top right insets illustrate the confidence intervals at the 68, 95, and 99\%  level. 
}
\label{plot_nc_lmbins}
\end{figure*}

Using the prescription described above, we computed  the stellar mass function of LBGs at $z\sim4$ and $5$. The results are shown in Figure \ref{plot_nc_lmbins} together with other measurements in the literature \citep{marchesini09,marchesini10,stark09,caputi11,gonzalez11,santini12}. We also indicate the number density measured in the GOODS-S and GOODS-N fields as small open symbols at each mass bin (circles and triangles). Our measurements at $z\sim4$ are in broad agreement with those of \citet[][blue open circles using set 7 in their Table 4]{marchesini09, marchesini10}, who measured the {\it total} stellar mass function of {\it all} galaxies at $3<z<4$. The fact that, at the massive end, their estimated number density is a factor of $2-3$ higher than ours is consistent with their observation that only $\approx30$\% of their sample at the very massive end would be selected as Lyman-break galaxies, while the rest is either passive or heavily obscured by dust. The measurements of \citet[][green downward triangles]{santini12} at $3.5<z<4.5$, mostly consistent with those of \citet{marchesini09}, are also in line with our data in their intermediate-mass bins. In their least massive bin $\approx10^{10}M_\odot$, their point is higher by a factor of 2, which may be due to cosmic variance as the bin includes only a handful of sources. We also compare our results with those of \citet[][cyan open and filled diamonds for $z\sim3.2$ and $z\sim3.9$, filled diamonds for $z\sim5$]{caputi11} after we corrected their masses by 0.16 dex. The correction is made to convert their masses (assuming Salpeter IMF) to ours (Chabrier IMF).  While at  $\log [M_{\rm{star}}/M_{\odot}] > 10.5$, their data are consistent with our estimates, they are systematically lower by more than a factor of 2 than our estimates  as well as those of \citet{marchesini09, marchesini10} and \citet{santini12} at lower masses at both redshift bins. While the origin of this discrepancy is unclear, possible causes include that 1) some of real high-redshift sources may have been missed by their photometric redshift selection, or 2) their incompleteness correction may have been underestimated.  Relatively large discrepancies present among similar studies based on the rest-frame optical selection \citep[e.g.,][]{fontana06,perez_gonzalez08,marchesini09,marchesini10,caputi11,santini12} may also be explained at least in part by  nonnegligible cosmic variance. Even the lowest-mass galaxies selected by these studies are amongst the most massive galaxies at high redshift, which are observed to be strongly clustered \citep[e.g.,][]{lin11}. 

We also compare our results with those that measured the SMF for star-forming galaxies. The results of \citet[][blue triangles]{stark09} are in god agreement with ours when we confine our analyses only to galaxies at $M_{1700}<-20$ as they did. A slight discrepancy at the massive end is likely due to our stringent selection, which resulted in removal of a few galaxies at the massive end, and possibly due to slightly different criteria applied to remove likely low-redshift interlopers. These ``cleaning'' processes affect the massive end the most, and clearly, the two measurements are comparable in  lower-mass bins at $\log [M_{\rm{star}}/M_\odot] <10.5$.  Finally, \citet[][orange upward triangles]{gonzalez11} found the low-mass end slope of the SMF to be $\alpha_M\approx-1.4$ in excellent agreement with our analyses. At $z\sim3.7$, their measurements are consistent with ours in their low-mass bins ($10^8M_\odot < M_{\rm{star}}<10^{9.5}M_\odot$) even though they went further down to below $10^8M_\odot$, whereas we chose not to go below $10^{8.7}M_\odot$. At $z\sim5$, their overall normalization is about 50\% higher than ours even though the estimated low-mass slope is still $\alpha=-1.39$ very close to ours $\alpha_M=-1.36$ (see discussions later). At the massive end, however, their data points are lower by more than a factor of 2 compared to our own results and those of \citet{stark09}. Because their sample is based on the WFC3 Early Release Science \citep[ERS;][]{windhorst11} data which covers about one third of the GOODS-S field (i.e., $\approx 15$\% of the total area covered by our study and that of \citet{stark09} who used both GOODS-N and GOODS-S), it is possible that their massive bins suffer more from cosmic variance than others. 

Based on our measurements, we determined the best-fit parameters when a Schechter function is assumed for the stellar mass function:
\begin{eqnarray}
\phi_M(M_{\rm{star}})&=&(\ln 10)\phi_M^*[10^{(M_{\rm{star}}-M^*_{\rm{star}})(1+\alpha_{\rm{SMF}})}] \nonumber \\
                                       && \times \exp[-10^{(M_{\rm{star}}-M^*_{\rm{star}})}]
\end{eqnarray}
Note that all parameters have subscripts ``M'' to distinguish from similar Schechter parameters for the UV luminosity function discussed previously.  To properly account for the measurement errors in mass, we use the probability distribution discussed in \S\ref{galsim} (also shown in Figure \ref{plot_sedfit_in_out} and \ref{plot_seds}) to convert the {\it intrinsic} SMF to the {\it observed} one: 
\begin{equation}
\phi_{\rm{M}}(M_{*,m}) = \Sigma_k \phi_{\rm{M}}(M_{*,k})~ p(M_{*,m}|M_{*,k})
\end{equation}
where $M_{*,k}$ is the true stellar mass of a galaxy and $M_{*,m}$ is the measured one. 
We use the chi-square minimization method to search for the best parameter set ($\alpha_{\rm{SMF}}$, $M^*_{\rm{star}}$). For a given parameter set ($\alpha_{\rm{SMF}}$, $M^*_{\rm{star}}$), we determine the normalization parameter $\phi_M^*$ such that the total number density of galaxies at $M_{\rm{star}}\geq M_{\rm{min}}$ equals the observed number density. The minimum stellar mass $M_{\rm{min}}$ is set to $10^{8.7} M_\odot$ and $10^9 M_\odot$ at $z\sim4$ and $5$, respectively for the reasons discussed in \S\ref{galsim}.  However, our results are insensitive to the minimum mass as long as it is set within the mass range over which the determination of stellar mass is relatively robust. 
On top right inset in Figure \ref{plot_nc_lmbins}, we show the confidence levels for the best-fit Schechter parameters, namely, the characteristic stellar mass $M^*_{\rm{star}}$ and low-mass-end slope $\alpha_{\rm{SMF}}$. 

At $z\sim4$, the best-fit value for the low-mass-end slope $\alpha_{\rm{SMF}}$ is $\approx-1.4$. Also at $z\sim5$,  the best-fit slope is $-1.4$ even though a large degeneracy between $M^*_{\rm{star}}$ and $\alpha_M$ (due to lack of constraints on the massive-end) makes a relatively steep slope also acceptable. The slope is much shallower than the faint-end slope of the UV luminosity function for the same galaxies, but is similar to that found from stellar-mass-selected studies \citep{fontana06, marchesini09}, and a recent study of LBGs based on a smaller sample by \citet{gonzalez11}. To visually illustrate this difference, we show in Figure \ref{plot_nc_lmbins} (dashed lines) the best-fit Schechter function when the slope $\alpha_{\rm{SMF}}$ is fixed to the faint-end slope of the UV luminosity function from \citet{bouwens07}, the values of which are indicated in Figure \ref{plot_nc_lmbins}. Clearly, a very steep low-mass end slope $\alpha_{\rm{SMF}}=-(1.7-1.8)$ does not provide a good fit to the observed stellar mass function even for the $z\sim5$ case. The shallow slope on the low-mass end seems to be a robust feature even considering somewhat large uncertainties for the normalization parameter as evidenced by a relatively large cosmic variance particularly at $z\sim5$. In Table \ref{tab:smf_parameters}, we summarize the best-fit Schechter parameters for both samples. For the $V_{606}$-band dropouts, we also provide the best-fit parameters when the characteristic mass $M_{\rm{star}}^*$ is forced not to exceed that at $z\sim4$. Such a treatment generally helps constrain the low-mass end slope better by reducing the large degeneracy between the two parameters. While it remains to be seen if it is indeed the case, the brightening of the characteristic luminosity $L_{UV}^*$ observed at high redshift \citep{bouwens07} likely supports such a scenario unless the mass-to-light ratios change significantly with redshift to reverse the effect. We also note that removing the least robust mass bin also makes the slope even shallower (albeit leaving larger overall uncertainties). Finally, we compare our measurements against recent predictions by \citet{guo11} using semi-analytic models implemented on the Millennium and Millennium II simulations at $z\sim4$ (shown as black lines on the left panel of Figure \ref{plot_nc_lmbins}). The dotted line indicates their predictions for the intrinsic SMF, while the solid line represents the values accounting for measurement errors (by convolving the intrinsic SMF assuming 0.3 dex scatter in stellar mass). On the massive end, our estimates are in excellent agreement with theirs. However, at the mass regime below the characteristic mass $\sim10^{10}M_\odot$, their model clearly overpredicts the abundance by a factor of $2-3$ compared to all the existing studies. Furthermore, the predicted slope for the low-mass end of the SMF is more in line with that of the UVLF shown in dashed line, much steeper than our measurements. 

Finally, we note that not all studies find the low-mass end slopes to be as shallow as our results as well as those of \citet{gonzalez11}. For example, \citet{kajisawa09} studied the galaxy SMF out to $z\sim3$, and reported that the low-mass end slopes become steeper with redshift, from $\alpha_M=-(1.2-1.3)$ at $z\sim0.7$ to $\alpha_M=-1.6$ at $z\sim3$. \citet{santini12} reported a similar steepening of the slope with redshift, from $\alpha_M=-1.44\pm0.03$ at $z\sim0.8$ to $-1.86\pm 0.16$ at $z\sim3$. For galaxies at $z\sim3.2$, $3.9$, and $4.7$, \citet{caputi11} found the low-mass end slope to be $\approx -1.8$ at all probed redshift bins. With the current observational constraints, these results are not necessarily contradictory to shallower slopes reported for  the SMFs of star-forming galaxies. Most studies based on mass-selected samples agree that the characteristic mass $M_{\rm{star}}^*$ lies well above $10^{11}M_\odot$ within a range of population synthesis models and assumptions (see all the references discussed above). On the other hand, the number density of star-forming galaxies at around $10^{11}M_\odot$  is observed to be very low, and consequently, the SMF measured only for star-forming galaxies tends to have a much lower characteristic mass ($10^{10}-10^{10.6}M_\odot$) in most studies. Such low characteristic mass is found independent of whether the masses are fit  for individual galaxies  \citep[such as this work and][]{gonzalez11} or by an average mass-to-light ratio was assumed \citep[e.g.,][]{reddy09,mclure09}. Hypothetically, if the fraction of star-forming (therefore, UV-visible) galaxies increases towards lower masses, then it is conceivable that the {\it true} mass function will have a high characteristic mass above $10^{11}M_\odot$ but will converge towards that of SMF at low masses where galaxies are dominated by star-forming galaxies. To test this possibility, we tried to fit our data to a Schechter function after forcing the characteristic mass to be greater than $10^{11}M_\odot$. When we only include our data points down to $\approx 3\times10^{9}M_\odot$, which is the lowest mass bin currently probed by deep surveys \citep{kajisawa09}, we find that the slope as steep as $-(1.8-1.9)$ can provide a good fit. However, once we include the lower mass bins, such a steep slope is  no longer acceptable. In other words, the steep slope found by mass-selected studies may be explained by a combination of the $M_{\rm{star}}^*-\alpha_M$ degeneracy (see inset of Figure \ref{plot_nc_lmbins} for the elongated shape of the contours) and the limited mass range currently probed by mass-selected samples, {\it if} the low-mass end is primarily dominated by star-forming galaxies. Deeper data are critically needed to extend the mass range and the fraction of star-forming galaxies at different masses, to help better understand the observed discrepancies between different studies.

\begin{deluxetable*}{ccccccc}
\tabletypesize{\footnotesize}
\tablecaption{The Schechter parameters for the stellar mass function of LBGs at $z\sim4$ and $\sim5$.\label{tab:smf_parameters}}
\tablehead{
& \multicolumn{3}{c}{$B_{435}$-dropouts ($z\sim4$)} & \multicolumn{3}{c}{$V_{606}$-dropouts ($z\sim5$)} \\
\colhead{SFH} & \colhead{$\log M_{\rm{star}} ^{*}$} & \colhead{$\phi_M^*$ (10$^{-4}$ Mpc$^{-3}$)} & \colhead{$\alpha_{\rm{SMF}}$} & \colhead{$\log M_{\rm{star}} ^{*}$} & \colhead{$\phi_M^*$ (10$^{-4}$ Mpc$^{-3}$)} & \colhead{$\alpha_{\rm{SMF}}$}}
\startdata
All SFHs\tablenotemark{a} & $10.35^{+0.30}_{-0.35}$ & $3.9^{+6.3}_{-2.4}$ & $-1.40^{+0.28}_{-0.20}$ & $10.45^{+1.05}_{-0.65}$ & $1.1^{+4.5}_{-1.0}$ & $-1.36^{+0.76}_{-0.48}$ \\
All SFHs\tablenotemark{b} & --- & --- & --- & $10.35_{-0.55}$ & $1.4^{+4.1}_{-0.6}$ & $-1.28^{+0.68}_{-0.28}$ \\
CSF\tablenotemark{a} &  $10.35^{+0.25}_{-0.30}$ & $3.7^{+5.6}_{-2.4}$ &  $-1.46^{+0.26}_{-0.22}$ & $10.50^{+1.00}_{-0.55}$ & $0.9^{+3.1}_{-0.9}$ &  $-1.42^{+0.52}_{-0.44}$\\
CSF\tablenotemark{b} &  --- &--- &  --- & $10.35_{-0.40}$ & $1.4^{+2.6}_{-0.5}$ &  $-1.32^{+0.42}_{-0.24}$\\
300 Myr\tablenotemark{a} &  $10.35^{+0.50}_{-0.40}$ & $3.5^{+7.7}_{-2.8}$ &  $-1.40^{+0.38}_{-0.28}$ & $11.20^{+0.30}_{-1.10}$ & $0.2^{+2.3}_{-0.1}$ &  $-1.60^{+0.50}_{-0.26}$\\
300 Myr\tablenotemark{b} &  --- & --- &  --- & $10.35_{-0.25}$ & $1.2^{+1.2}_{-0.4}$ &  $-1.34^{+0.24}_{-0.20}$\\
100 Myr\tablenotemark{a} &  $10.15^{+0.55}_{-0.35}$ & $4.0^{+6.8}_{-3.2}$ &  $-1.40^{+0.34}_{-0.28}$ &  $10.50^{+1.00}_{-0.55}$ & $0.6^{+2.4}_{-0.5}$ &  $-1.50^{+0.50}_{-0.46}$ \\
100 Myr\tablenotemark{b} & --- & --- &  --- &  $10.35_{-0.40}$ & $1.0^{+2.0}_{-0.5}$ &  $-1.40^{+0.40}_{-0.32}$ \\
\enddata
\tablenotetext{a}{The characteristic mass $\log M_{\rm{star}} ^{*}$ is allowed to vary during the fit in the mass range $9\leq \log M_{\rm{star}} ^{*} \leq 11.5$}
\tablenotetext{b}{The characteristic mass $\log M_{\rm{star}} ^{*}$ is allowed to vary only below $M_{\rm{star}} ^{*}=10.35$, the best-fit value for the $B_{435}$-band dropouts }
\end{deluxetable*}

\section{Discussions}\label{discussions}
We have measured the SED properties of LBGs at $z\sim4$ and $5$, two of the largest photometric samples ever assembled consisting of 3088 and 987 galaxies, respectively. The use of TFIT, which  can very robustly measure galaxy light within a fixed physical aperture over data with a wide range of PSF sizes, and a large set of galaxy simulations to quantify measurement errors and systematics of TFIT photometry at different flux levels, have allowed us to make the most robust measurements to date of galaxy properties, such as stellar mass, age, and UV luminosity, for a large sample, and carry out a large statistical study of high-redshift star-forming galaxy population as a whole. Based on these measurements, we concluded that, 1) there is a significant {\it intrinsic} scatter to the correlation between the observed UV luminosity and stellar mass, 2) the number counts of galaxies in the bins of UV luminosity strongly support the previous studies \citep{bouwens07,reddy09}, that the UV luminosity function at these redshifts increases steeply $\alpha_{\rm{UVLF}}\approx -(1.7-1.8)$ towards the faint end, and finally, 3) the number density of the same galaxies in the bins of stellar mass increases rather mildly $\alpha_{\rm{SMF}}\approx -(1.3-1.4)$ compared to the UV luminosity function of the same galaxies. In what follows, we discuss these findings in further detail and their possible implications on how these UV-selected galaxies assemble their masses over cosmic time. 

 \subsection{The $M_{1700}$-$M_{\rm{star}}$ Scaling Relation}\label{M1700_Mstar}
 The locations of galaxies on the SFR-$M_{\rm{star}}$ plane indicate typical star formation histories of the galaxy population in question. If most galaxies have been assembling their mass smoothly over a long period of time, one would expect them to form a tight ``main-sequence'' on the SFR-$M_{\rm{star}}$ plane as stellar mass is, to the zeroth order, the time integral of the past star formation history. Indeed, a relatively tight main-sequence has been observed for star-forming galaxies in the local universe \citep{noeske07}, and for massive star-forming galaxies at $z=1$ and $2$ \citep{elbaz07,elbaz11,daddi07a,rodighiero11}. The observed scatter around the main sequence is $0.25$ dex in all cases.  For galaxy population at $z>2$, however, the trend is still unclear. While a similar (positive) correlation between star formation rates and stellar mass exists \citep[][from stacking]{pannella09, lee11}, many observations reported much larger scatter \citep[$\sigma\gtrsim0.46$ dex;][]{reddy06a, mannucci09, magdis10}. 

As shown in Figure \ref{2d_nc},  our results suggest that there is a significant scatter in the correlation between UV luminosity ($M_{1700}$) and stellar mass. It is evident at both redshifts that at a fixed stellar mass, there is a non-negligible fraction of galaxies that extend down to nearly 2 magnitudes fainter than the ``main-sequence''. It is difficult to use these observations to estimate the true extent of this faint tail because our samples are UV-selected, and thus are not complete in mass. It is worth noting that the UV selection misses a large number of very faint galaxies ($M_{1700}>-19.5$) as the completeness of the data decreases at the faint end. If we correct for the missing galaxies, the faint-end tail is even more significant than that observed here. Furthermore, as the UV selection systematically misses red galaxies with heavily obscured star formation, inclusion of such a population would further increase the scatter. 

The main limitation of interpreting the observed $M_{1700}$-$M_{\rm{star}}$ scaling relation is the fact that UV light from these galaxies are systematically dimmed by internal dust \citep{calzettietal97, calzetti00}. At a fixed UV luminosity, the distribution of UV extinction can vary significantly \citep[e.g.,][]{bouwens09, reddy10}. For example, the $1\sigma$ scatter in the color excess $E(B-V)$ of 0.1 (or extinction at 1700\AA, $A_{1700}$, of 1.0 mag) means that galaxies at a fixed UV luminosity can have SFRs different up to a factor of $6$ (from $-1\sigma$ to $1\sigma$). In turn, the intrinsic distribution of UV extinction can cause a significant scatter in the observed $M_{1700}$-$M_{\rm{star}}$ scaling relation, even if the intrinsic relation between SFR and $M_{\rm{star}}$ is tight. To account for this effect, we use the UV-slope measurements of \citet{bouwens09}.  
The main reason for using \citet{bouwens09} measurements instead of our own is their inclusion of the Hubble Ultra Deep Field and HUDF Parallel Fields in the analyses. These data not only provide higher signal-to-noise measurements of colors for sources detected in the GOODS data, but also allows to define a larger sample of UV-faint sources form which the color distribution is computed. 
Assuming the \citet{calzetti00} extinction curve, the intrinsic scatter in the UV continuum slope $\beta$ (their Table 4) translates into the intrinsic scatter of $0.7-0.9$ mag in the extinction at 1700 \AA, or $A_{1700}$. A more recent result by \citet{castellano11} also found a similar scatter. As for the observed $M_{1700}$-$M_{\rm{star}}$ relation, we correct it for the selection completeness we estimated from our simulations. For example, at $M_{1700}=-21 ~(-20, -19)$, the completeness is 89 (72, 34)\%. Because our sample is selected based upon the UV colors and flux limit, it is fair to assume that at a fixed stellar mass correcting for the UV completeness   is appropriate. In other words, the distribution of stellar mass at a fixed UV luminosity has no bearing on altering the intrinsic distribution we attempt to recover here. 

\begin{figure}[t]
\epsscale{1.0}
\plotone{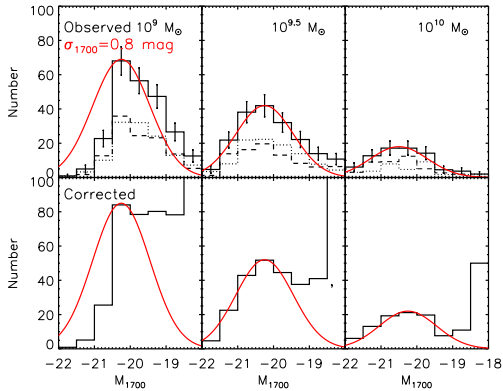}
\caption[plot_MUV_distribution]{
The distribution of absolute UV magnitudes in stellar mass bins of $M_{\rm{star}}=10^9, 10^{9.5},10^{10}M_\odot$ (from left). In upper panels, we show the observed distribution of UV magnitudes for all galaxies (solid histogram) and for galaxies in each field (dashed and dotted histogram) in the respective bins. The error bars indicate Poisson errors. The smooth function represents a normal distribution with standard deviation of $\sigma=0.8$ mag. The scatter is chosen to match the UV slope measurements by \citet{bouwens09}, and its correlation with extinction \citep{meurer99}. If dust extinction alone is responsible for the observed spread of the observed UV luminosity, then the function would closely reflect the observations.  In lower panels, we show the same distribution but corrected for the selection incompleteness. 
}
\label{plot_MUV_distribution}
\end{figure}

In Figure \ref{plot_MUV_distribution}, we show the observed UV luminosity distribution of galaxies at $z\sim4$ at three different stellar mass bins, $M_{\rm{star}}=10^{9}, 10^{9.5}, 10^{10}M_\odot$. Histograms represent the distribution of absolute UV magnitudes for all galaxies (solid), and in individual fields (dashed and dotted for GOODS-S and GOODS-N, respectively). The smooth function in each panel approximates the emergent distribution of $M_{1700}$ when there is a perfect correlation between SFR intrinsic UV luminosity, $L_{1700}$, and stellar mass  and dust extinction is solely responsible for the spread in the observed UV luminosity.
All functions have a standard deviation of $\sigma=0.8$ mag as discussed above, and are renormalized to match the peak of the observed distribution. We find that there is an excess of UV-faint galaxies at least in the lowest mass bin. The observed distribution (top left) already suggests the excess although marginally. However, when the distribution is corrected for the selection incompleteness, the existence of such excess becomes more apparent. At $M_{1700}\sim-19.5$, the number of galaxies is higher by at least a factor of 2 when corrected for the selection effect. The selection completeness at this luminosity is still relatively high ($\approx$50\%), and thus our estimate should be robust. Towards the massive end, the discrepancy between the observed distribution and that expected from UV extinction alone appears to lessen in both observed and corrected distribution. In particular, at the highest-mass bin, the observed distribution is in excellent agreement with the smooth function (except at the faintest end, where the completeness level is very low, and probably not very robust). The implication is that the dust-corrected UV luminosity (directly proportional to SFR) and stellar mass are very tightly correlated at the massive end, while there is a significant scatter in the relation at the low-mass end. 

\subsection{Inferences from the Shape of the UV Luminosity Function and Stellar Mass Function}\label{double_power_law}
Now we turn our attention to the stellar mass function of LBGs. In \S\ref{UVLF}, we showed that the number counts of galaxies in the UV luminosity bin are consistent with a steep faint-end slope ($\alpha_{\rm{UVLF}}\approx -(1.7-1.8)$), in excellent agreement with other LF measures of high-redshift star-forming galaxies in the literature \citep{bouwens07,reddy09}. Interestingly, when we estimated the stellar mass function of the {\it same} galaxies (\S\ref{SMF}), we found that the low-mass end slope of the mass function is relatively shallow, $\alpha_{\rm{SMF}}=-(1.3-1.4)$, compared to the faint-end slope of the UV luminosity function (see Figure \ref{plot_nc_lmbins}). The overall $M_{1700}-M_{\rm{\star}}$ correlation discussed in \S\ref{M1700_Mstar} means that the majority of the galaxies on the low-mass end of the stellar mass function mainly populate the faint end of the UV luminosity function. The implication is that a substantial fraction of the galaxies on the faint end of the UVLF (which is very steep) have stellar mass lower than what the sensitivities of our data allow (i.e., $M_{\rm{star}}\sim10^{8.7}M_\odot$ and $10^{9}M_\odot$ at $z\sim4$ and $5$, respectively; see \S\ref{galsim}). 

The relatively shallow low-mass end slope of the SMF has important implications for typical star formation history of galaxies in our sample. If most galaxies have been forming stars in the past at comparable rates as when observed, then, to the zeroth order, the stellar mass function of LBGs should have a similar shape as the UV LF  as $M_{\rm{star}}\propto$ SFR. Introducing scatter to this relation increases the overall normalization of the function, but does not alter the overall shape or the low-mass end slope as long as the scatter is roughly constant. What is needed to effectively map galaxies in the UV LF to that in the SMF is to assume luminosity-dependent stellar-mass-to-UV-light ratios. This may mean either the median ratio $M_{\rm{star}}/L_{1700}$ is systematically lower for galaxies on the faint end of the UVLF than more UV-luminous ones or the  $M_{\rm{star}}/L_{1700}$ distribution has a long tail towards the low  $M_{\rm{star}}/L_{1700}$ end. Both scenarios would effectively result in a large number of UV-faint galaxies to have stellar masses lower than the minimum stellar mass ($\approx 10^{8.7}M_\odot$) that we deem reliable. 

In order to make more quantitative statements on these interpretations, we consider toy models in which the $M_{\rm{star}}$-$M_{1700}$ scaling law is described by a mean and scatter.  We begin by taking the UV LF, assign stellar masses to the  UV-selected galaxies according to the scaling law, and thereby compute the stellar mass function of the same galaxies. This can be done by a simple integral:
\begin{equation}
\phi_M(\log M_{\rm{star}})=\int_0^\infty \phi(M_{1700}) \mathcal{P}(\log M_{\rm{star}}|M_{1700}) dM_{1700}
\end{equation}
where $M_{\rm{star}}$ is in units of $M_\odot$ and $ \mathcal{P}(\log M_{\rm{star}}|M_{1700})$ is a probability distribution of stellar mass at a fixed UV luminosity, which we assume to be a normal distribution. At the bright end ($M_{1700}\lesssim -(20-21)$), we fix the mean scaling law (where the probability distribution peaks) to what we observe in Figure \ref{2d_nc}:
\begin{equation}\label{scaling_law}
\log M_{\rm{star}} = \gamma_{\rm{bright}}~M_{1700}+1.2
\end{equation}
We assume $\gamma_{\rm{bright}}=-0.415$, the value that is consistent with the observed scaling relation at the bright end \citep[][also see Figure \ref{2d_nc}]{stark09, lee11}. A slope close to $-0.4$ means that the UV luminosity, $L_{1700}$, scales linearly with stellar mass. Furthermore, we assume a constant scatter of $0.3$ dex around the mean scaling law. While the scatter has never been measured robustly for these galaxies, \citet{lee11} concluded that the $M_{1700}-M_{\rm{star}}$ correlation must be rather tight on the bright end based on the stacking analysis of a large sample of $L\gtrsim L^*$ ($M_{1700}<-21.5$) galaxies at $z\sim4$. In any case, adding more scatter does little other than increasing the overall normalization of the SMF. This is because a larger number of galaxies on the faint end of the UVLF scatters into the massive end of the SMF than UV-luminous galaxies scatter into the low-mass end (similar to the ``Malmquist bias''). As scatter does not affect the low-mass end slope, and we have no means to robustly measure this scatter observationally for the UV-faint galaxies, we set the scatter to be 0.3 dex throughout this discussion. 

\begin{figure*}[t]
\epsscale{0.9}
\plotone{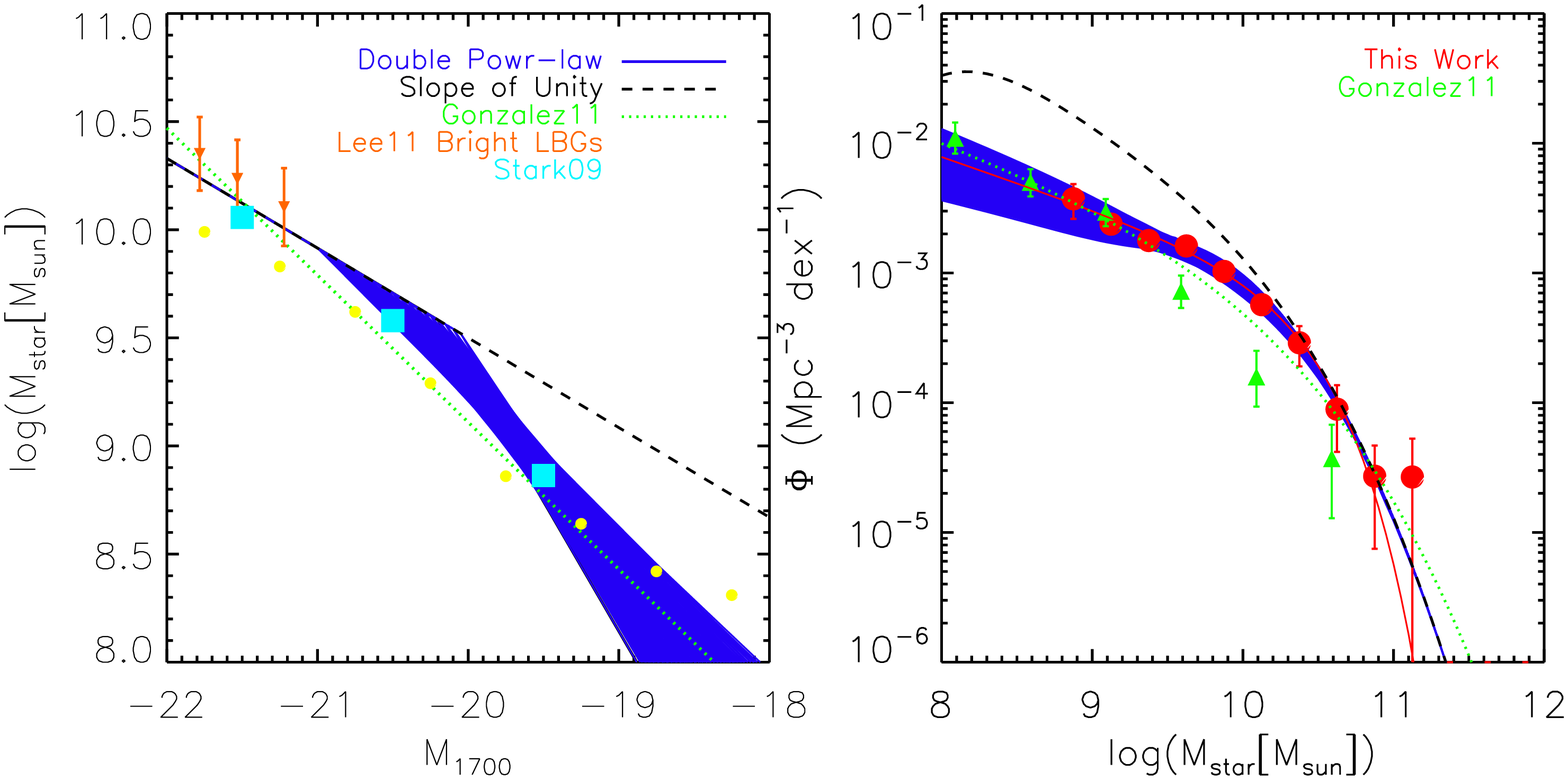}
\caption[plot_double_power_law]{
The  inferred $M_{\rm{star}}$-$M_{1700}$ scaling law of LBGs at $z\sim4$ are shown together with the observational measures \citep[][cyan squares and downward triangles, respectively]{stark09, lee11} (see \S \ref{double_power_law}). We also show the average stellar mass at a fixed UV luminosity directly determined from our measurements (yellow circles; see Figure \ref{2d_nc}). The $M_{\rm{star}}$-$M_{1700}$ scaling law is modeled as a double power law, and the acceptable models that reproduce the observed stellar mass function (right) are shown in blue. 
A single power-law model with the slope of unity well describes the observed $M_{\rm{star}}$-$M_{1700}$ relation on the bright end \citep[][downward triangles]{lee11}, but produces too many low-mass galaxies as the low-mass end slope is steep, $\alpha_{\rm{SMF}}=-(1.7-1.8)$, as shown in dashed lines on both panels. 
A model in which mass-to-light ratios decrease significantly towards the faint end provides a better description of the observed number counts in stellar mass bins, such as a double power-law (blue swaths) or a steeper slope \citep[shown in green as measured by][]{gonzalez11}. A steeper decline of stellar mass towards the faint end of the UVLF implies that these galaxies have not sustained their star formation at the rates seen at the time of observations. 
}
\label{plot_double_power_law}
\end{figure*}

First, we compute the SMF using the scaling law observed for the relatively UV-luminous galaxies (Equation \ref{scaling_law}). The results are shown in Figure \ref{plot_double_power_law} (dashed line). Clearly, the low-mass end slope in this case is much steeper than the observed one. This is expected because the assumed scaling law implies that the UV luminosity and stellar mass scale linearly ($M_{\rm{star}} \propto L_{1700}$), and thus the low-mass end slope of the SMF should have the identical functional shape as the UVLF. Based on the fact that the assumed scaling law reproduces the bright end of the SMF, and that it is determined from the observations where the measurements have the least uncertainties, we determine that the single power-law mapping between the observed UV luminosity and stellar mass does not provide a good description of our observations. 

Next, we try a double power-law model in which the scaling law is fixed to Equation \ref{scaling_law} at the bright end, but the slope changes to a different value at the faint end. This approach is a two-parameter model (the luminosity $M_{\rm{break}}$ at which the break takes place, and the power-law slope at the faint end $\gamma_{\rm{faint}}$). Over a wide range of $M_{\rm{break}}$ and $\gamma_{\rm{faint}}$, we computed the SMFs to find the best scaling law that describes the observed SMF. In Figure \ref{plot_double_power_law}, we illustrate the range of the scaling laws and their corresponding SMFs which return the reduced chi-square of $\chi^2_r \leq1.2$ (68\% confidence level). The best-fit model is $M_{\rm{break}}=-20.45$ and $\gamma_{\rm{faint}}=-0.84$. Clearly, all the acceptable scaling laws have the power-law slope $\gamma_{\rm{faint}}\leq-0.66$, much steeper than that on the bright end ($\gamma_{\rm{bright}}\approx -0.4$). 

Alternatively, it is possible that the scaling law has a steeper but single slope as suggested by \citet{gonzalez11}. In Figure \ref{plot_double_power_law}, we also show their best-fit scaling law ($\log M_{\rm{star}} = -0.68M_{1700}-4.49$ with 0.5 dex scatter around the mean) and corresponding SMF prediction in dotted lines. Indeed, their scaling law is consistent with the current observations in most luminosities \citep[][filled squares and downward triangles, respectively]{stark09, lee11} and is in good agreement with our estimation (blue swaths) assuming double power-laws. It also reproduces well the shallow slope of the galaxy SMF as measured independently by this work and theirs. The disagreement between the model SMF (dotted line) and their data (filled triangles in the right panel of Figure \ref{plot_double_power_law}) can likely be explained by the fact that they constructed the SMF using individual measurements from their data and not scaling by the mean mass-to-light ratios as we did here.  Their scaling law slightly under-predicts the galaxy space density at the intermediate masses ($\log [M_{\rm{star}}/M_\odot]=9.5-10.5$).  More data are likely needed to improve the accuracies of measurements in the relevant mass ranges. Despite minor discrepancies, the two studies clearly agree in that the observed mass-to-light ratios is luminosity-dependent such that it is considerably lower by a factor of 2-4 for sub-$L^*$ galaxies compared to UV-bright galaxies.  


\subsection{The Impact of Luminosity-Dependent Extinction}\label{impact_of_extinction}
As discussed in \S\ref{double_power_law}, the difference in the observed slopes between the UVLF and SMF of the same galaxies seems robust. As shown in Figure \ref{plot_double_power_law}, if galaxies obey a single power-law (dashed lines) on the observed $M_{1700}-M_{\rm{star}}$ plane, the abundance of low-mass ($\lesssim 10^{9.5}M_\odot$) galaxies would be overpredicted by a factor of several. However, the observed difference may not entirely be attributed to changing mass-to-light ratios {\it if} internal dust extinction strongly depends on UV luminosity. If the extinction systematically increases with UV luminosity, the space density of UV-bright galaxies  would be boosted more substantially than that of UV-faint galaxies after correcting for dust. As a result, the ``dust-corrected'' UV luminosity function (intrinsic UVLF, hereafter) would have a shallower faint-end slope than the observed UVLF.  Hence, it is important to quantify how much of this effect would be needed to explain the observed   $M_{1700}-M_{\rm{star}}$  scaling law without invoking changing mass-to-light ratios. 

Unfortunately, existing works are somewhat conflicting on this topic at present ranging from a relatively strong dependence \citep{bouwens09, bouwens11b, lee11, castellano11} to little or no dependence \citep{ouchi04a, dunlop11, finkelstein11} at $z\gtrsim4$. The discrepancies among different works likely lie in the limited knowledge of the precise redshifts for the samples, measurement errors arising from the use of a limited number of photometric bands (often one or two colors), and accounting for the selection effects of high-redshift samples that may mimic the trend, or quite possibly, a combination of these effects. Obviously, the strongest luminosity-dependent trend for extinction will have the most dramatic impact on the interpretations of our results. Therefore, we explore this possibility by using the most recent results of \citet{bouwens11b}, who reported a clear trend of luminosity-dependent extinction.  

We use their updated best-fit formula given by \citet{bouwens11b}, $\beta=-0.10(M_{\rm{UV}}+19.5)-1.98$  and the $1\sigma$ scatter of $\sigma_\beta=0.3$ \citep{bouwens09}. 
Assuming the correlation between the extinction ($A_{1600}$)  and UV slope ($\beta$) observed for local starburst galaxies \citep[][$A_{1600}=2.31\beta+4.85$]{calzetti00}, we estimate the effect of luminosity-dependent extinction on the shape of the UVLF.  We start with a Schechter function for the intrinsic UVLF, then apply luminosity-dependent dust extinction to repopulate galaxies on the observed UV luminosity space, and compute the observed UVLF. The Schechter parameters for the intrinsic UVLF were varied until we reproduced the observed UVLF at $z\sim4$ \citep{bouwens07}. The use of a Schechter function is not essential to our results as long as the faint-end is described as a power-law. The best-fit faint-end slope for the intrinsic UVLF is $-1.62$, somewhat shallower than that of the observed UVLF of $-1.76$. Assuming a slightly different correlation from \citet{meurer99} -- $A_{1600}=1.99\beta+4.43$ -- does not change the slope for more than $\Delta \alpha=0.01$. 

As expected, the resultant faint-end slope of the intrinsic UVLF is shallower than that observed if there is a positive correlation between extinction and UV luminosity. However, it seems that the ``flattening'' that occurs as a result of luminosity-dependent extinction is relatively mild even for the strongest trend reported so far, and thus is insufficient to explain the observed shallow slope $-(1.3-1.4)$ for the SMF at the $1.5\sigma$ level (the discrepancy is at the $\approx2\sigma$ level in the absence of such a trend). Hence, we conclude that changing mass-to-light ratios are indeed needed to explain the observed shape of the stellar mass function even if extinction is luminosity-dependent. Careful analyses on this topic will shed further light on more proper interpretations of our results.

\subsection{The Origin of the Changing Mass-to-Light Ratios}\label{origins}
In what follows, we consider mainly three physical scenarios that might have caused the observed scaling law to deviate from a single power-law. 
The first scenario we explore is that more UV-luminous galaxies have, on average, higher formation redshifts than UV-fainter ones, while {\it all} galaxies have sustained their star formation at relatively constant levels as observed. This allows us to reproduce the observed SMF while maintaining that most galaxies have formed over a long smooth star formation history. Such a picture is qualitatively in line with theoretical expectations that more massive galaxies formed earlier in more massive halos (``hierarchical formation''). Furthermore, smooth formation histories for average galaxies are favored by numerical simulations, as cold gas is accreted efficiently via filamentary structures without shock-heating \citep{keres05,keres09,dekel09}. However, this picture appears to directly contradict two observations. First, ages of galaxies would be luminosity-dependent in this scenario. Reading off the left panel of Figure \ref{plot_double_power_law}, galaxies with UV luminosity  $M_{1700}=-19.5$ has stellar mass of roughly one third (blue swath) of what they would have amassed if the star formation varies smoothly over time\footnote{We note that {\it any} star formation history that varies smoothly over time would produce a slope of unity, including a constant SFH, exponentially or linearly increasing/declining models unless galaxy ages (or formation redshifts) change with stellar mass} (dashed line). In other words, those galaxies should be on average 3 times younger than those at  $M_{1700}=-21$. 
Instead, we find that all galaxies have similar ages ranging over $200-400$ Myr when a constant star formation history is assumed, in agreement with \citet{stark09}.  However, we note that it would be hard to constrain population ages younger than 200 Myr as the Balmer break would be weak at the $0.1-0.2$ mag level. With the current depth of the available data (mainly limited by the near-infrared data from the ground), large photometric errors can easily wash away the trend, even if it existed. Second, perhaps more convincing observational constraint comes from the clustering studies. \citet{lee09} found that, based on the measurement of correlation function of identical samples, the observed small-scale clustering is inconsistent with halo clustering if the same number density is imposed (i.e., all halos above a mass threshold hosts a visible galaxy).
They argued that if only a fraction ($15-40$\%) of halos hosts galaxies at a given time, the shape of the correlation function for galaxies and halos would be well-matched. Because our first scenario posits that galaxies have long continuous formation histories, it directly contradicts the short duty cycle implied by the clustering studies by \citet{lee09}. 

The second possibility is that the formation redshifts of galaxies depend on luminosity, but the average star formation rates of galaxies also rise with time. Rising star formation history has been suggested  based on several high-redshift observations. These include the arguments that galaxies at high redshift appear to be very young while forming a relatively tight sequence on the SFR-$M_{\rm{star}}$ plane, which apparently contradict one another unless rising star formation history is invoked \citep{renzini09,maraston10,lee11}. Another compelling argument is that the median star formation rate at a fixed comoving density rises with redshift at least for the most UV-luminous galaxies \citep{papovich11}. Recent numerical simulations also support this view \citep[e.g.,][]{finlator11}. 
While a rising SFH would make galaxies appear young regardless of their formation redshifts,  the second scenario also predicts a long duty cycle, and thus contradicts the clustering constraints. 

Finally, we consider a scenario in which the average star formation history is more episodic. One immediate feature would be that galaxies are not UV-luminous at all times, and therefore only a fraction of galaxies would be observed at a given time. The exact fraction would be determined by how long UV-visible SF phase is sustained during the observed sampling of cosmic time. Furthermore, one would observe a substantial scatter on the SFR-$M_{\rm{star}}$ plane and $M_{\rm{star}}$-$M_{1700}$ plane (independent of any variations caused by a dispersion in extinction), as some galaxies would be observed at the rise or decline of their SF episode and thus lie far outside of the ``SF main sequence''. The episodic SFH scenario, however, does not preclude the existence of the main sequence albeit with larger scatter than that expected from continuous SF scenarios. In this case, the main sequence would represent how the average strength of each episode depends on halo mass. In fact, the halo-mass-dependent UV star formation scenario is supported by clustering studies that the correlation length (i.e., the clustering strength) increases with UV luminosity, implying that more UV-luminous galaxies are, on average, hosted by more massive halos \citep{GD01, ouchi04b, adelberger05, kashikawa06, lee06, lee09, hildebrandt07,hildebrandt09}. Such requirements can be met if more massive halos accrete, on average, a larger amount of cold gas to fuel star formation episodes. In such a picture, a rising star formation history is a direct result of the fact that  the host halo masses and thus possibly cold gas accretion therein increase with cosmic time. In addition, the episodic SF scenario would be able to explain young ages observed at high redshift as the galaxy light is dominated by a recently formed stellar population, even if there exist older generations of stars formed from earlier episodes. 

It is worth noting that, in this interpretation, the observed $M_{\rm{star}}$-$M_{1700}$ relation can be used to estimate the median SF duty cycle (or the duration of typical SF episode).  Because the unit power-law slope of the $M_{\rm{star}}$-$M_{1700}$ relation is achieved when SF varies smoothly over time, the deviation from it can immediately be interpreted as due to a non-unity duty cycle. In this regard, the left panel of Figure \ref{plot_double_power_law} can provide a useful diagnostic to estimate the SF duty cycle. To the zeroth order, the duty cycle should be a ratio of the stellar mass that galaxies should have amassed  if SF was continuous at the observed level (dashed line indicating the unit power-law slope) to the observed stellar mass (blue swath). Interestingly, the duty cycle of our $z\sim4$ sample computed using this method is $\approx20-40$\% (evaluated at $M_{1700}\approx -19.4$ AB, roughly corresponding to the survey depth), in good agreement with that estimated from our clustering study \citep{lee09}.  

One interesting implication of this interpretation is that the SF duty cycle changes with luminosity or halo mass. At the UV-luminous regime ($M_{1700}\lesssim-21$), star formation proceeds rather smoothly with long duty cycles and possibly at increasing rates. This is in line with other studies of UV-luminous galaxies \citep{papovich11, lee11} at these redshifts. On the other hand, at the UV-faint end ($M_{1700}\gtrsim-20$), the dominant mode of SF becomes progressively shorter and thus more ``episodic''.  This may be due to more sporadic replenishment of cold gas in relatively low-mass halos \citep{lee09}, which likely host these UV-faint galaxies, or due to increasingly more efficient feedback mechanism which can easily shut down star formation by driving away available cold gas \citep{finlator11}. Whatever the main driver may be, the differential evolution of SF in dark matter halos suggested by this scenario presents a nontrivial test to theoretical models of galaxy formation. 

\subsection{Star Formation Duty Cycle vs. Invisible Halos Scenario from \citet{finlator11}}
In light of our findings discussed above, it is interesting to consider an alternate interpretation to the short SF duty cycle scenario that explains the observations at high redshift. The two main observations include uniformly young population ages \citep{stark09}, and a relatively low halo occupancy implied by the clustering study \citep{lee09}. The former implies that galaxies observed at higher redshift (e.g., $z\sim6$) cannot be progenitors of those observed at lower redshift (e.g., $z\sim4$) because their population ages are similar. On the other hand, the clustering study implies that only a fraction ($15-40$\%) of all halos host a visible star-forming galaxy at a given point in time. Both can be explained naturally if  UV-visible star-formation does not continue for the Hubble time, but rather has a typical timescale of $\approx300$ Myr. 

Another interpretation put forward by \citet{finlator11}  is that there is a significant scatter in the baryonic mass and halo mass relation in low-mass halos. Motivated by hydrodynamic simulations, they predict that galaxies in a significant fraction of low-mass halos will be affected more severely by strong outflows \citep[``momentum-driven winds'';][]{dave06} which suppress star formation therein. In turn, these galaxies would have much lower stellar mass than others hosted in halos of similar masses. As a result, galaxies hosted by low-mass halos would span a broad range of star formation rates (and stellar masses), and the observed galaxies only represent the top end of this distribution (i.e., galaxies whose star formation is least suppressed by feedback). This scenario can also naturally explain a low halo occupancy number similar to that measured by  \citet[][$0.15-0.40$]{lee09} for LBGs, as a large fraction of halos will remain ``dark'' out of observational reach. They further predicted that the observed UV-selected galaxies will go on to have a smoothly rising star formation history with a duty cycle of unity, while others will remain undetected by observations., which explains the young ages (their stellar population is always dominated by recently formed stars). 

One caveat in their scenario is that  the observed galaxies would form a tight sequence on the SFR-$M_{\rm{star}}$ plane (see their Figure 10) as their SFH is smoothly rising with time. One direct result is that the stellar mass function of LBGs should still closely mirror the UV luminosity function and thus the low-mass end  slope of the SMF is close to $\alpha_{\rm{UVLF}}\sim-2.0$, a typical value for the faint-end slope predicted by simulations \citep[e.g.,][]{finlator06,choi10}.  A very steep  SMF slope is in direct conflict with our findings and those of \citet{gonzalez11}, that the stellar mass function rises considerably more shallowly than the UVLF of the same galaxies and that the stellar mass to UV light ratio indeed changes with luminosity. More careful comparison of the observed $\mathcal{M_{\rm{star}}/L_{\rm{UV}}}$ with the predictions on a larger sample will provide a useful test for the validity of such a scenario.

\section{Summary}
Using one of the deepest multi-wavelength data sets on two independent fields, we have investigated the statistical properties of star-forming galaxies (LBGs) at $z\sim4$ and $5$, namely, their number counts as a function of UV luminosity and stellar mass, as well as how these  two quantities are related to each other. These statistics have important implications as to the average star formation history of these galaxies as they represent the ongoing star formation and the time integral of the past star formation, respectively. Based on our analyses, we conclude:\\

1. The galaxy counts in the rest-frame UV magnitudes (at 1700\AA) are consistent with a steep faint-end slope, $\alpha\approx-(1.7-1.8)$, of the UV luminosity function, in good agreement with several existing studies at high redshift \citep[][see \S\ref{UVLF} and Figure \ref{plot_nc_mbins}]{reddy09, bouwens07}. \\

2. Based on the locations on the $M_{1700}$-$M_{\rm{star}}$ plane populated by galaxies in our samples, we find a broad correlation between stellar mass and UV luminosity, such that more UV-luminous galaxies are, on average, also more massive (\S\ref{M1700_Mstar}; see Figure \ref{2d_nc}). However, at both $z\sim4$ and $5$, the correlation has a substantial intrinsic scatter, in particular, for UV-faint galaxies. This is evidenced by the fact that there is a non-negligible number of UV-faint but massive galaxies that are present in our samples. Roughly 35 (20)\% of the massive galaxies ($M_{\rm{star}}\geq10^{9.5}M_\odot$) are UV-faint ($M_{UV}>-20$) at $z\sim4$ (5). In contrast to our findings for UV-faint galaxies, we also report that the region of the low stellar mass and high UV luminosity is largely devoid of galaxies, suggesting that the majority of the UV-luminous galaxies may have somewhat more extended star formation history (for at least several hundred million years) than their UV-fainter counterparts.  \\

3. While the current data do not allow us to directly quantify the {\it intrinsic} scatter of the SFR-$M_{\rm{star}}$ scaling law, the distribution of the UV magnitudes at a fixed stellar mass strongly suggests that there exists a significant intrinsic scatter at least on the low-mass end (Figure \ref{plot_MUV_distribution}). Considering the adopted UV color selection which systematically misses highly dust-obscured systems, the true distribution is likely even wider than that estimated in our analyses. Better sensitivities in the near-infrared (sampling the UV slope $2000-3500$\AA) and mid-infrared (sampling longward of the Balmer/4000 \AA\ break) are needed to make more precise measurements of the intrinsic SFR-$M_{\rm{star}}$ scaling law. \\

4. We make the statistically robust estimates of the stellar mass functions for LBGs at $z\sim4$ and $5$ (\S\ref{SMF}). Our measurements suggest that the low-mass end slope of the SMF is  $\alpha_{\rm{SMF}}\approx-(1.3-1.4)$, and thus is not as steep as that of the UVLF of the same galaxies at both redshifts (Figure \ref{plot_nc_lmbins}), in agreement with a recent study by \citet{gonzalez11} based on a smaller sample. The direct implication is that a large fraction of the UV-selected galaxies are not massive enough, and therefore are too faint in their rest-frame optical bands, to be detected in the current IRAC data. The broad $M_{1700}$-$M_{\rm{star}}$ correlation observed for these galaxies implies that those ``missing'' galaxies with stellar mass $M_{\rm{star}} \lesssim 10^{8.7}M_\odot$ are mostly UV-faint. In a scenario in which most galaxies have a relatively continuous star formation history, these galaxies would have accumulated more stellar mass (by a factor of several) than observed, and as a result, the SMF would more closely mirror that of the UVLF of the same galaxies. Hence, our results favor a more episodic formation history in which star formation rates of galaxies largely fluctuate over cosmic time. The duty cycle  inferred from our SMF measurements is $20-40$\% (\S\ref{origins}), in good agreement with that implied from a clustering study by \citet{lee09}. \\

5. Using a simple toy model, we demonstrate that the stellar mass to UV light ratio should decrease systematically for galaxies with $M_{1700} \gtrsim-20.5$ (\S\ref{double_power_law}; also see Figure \ref{plot_double_power_law}) in order to reproduce the observed low-mass end slope of the SMF.  Assuming a roughly constant mass-to-light ratios for all galaxies would result in significant overestimation for the abundance of low-mass galaxies, in direct violation of  the current SMF measurements. We explore a possible impact of the luminosity-dependent dust extinction on the interpretations of our results using the strongest trend reported to date, and conclude that it is likely insufficient to explain the observed shape of the SMF, although such a trend would imply a milder decline of the mass-to-light ratios towards faint UV luminosities than that expected in the absence of the luminosity-dependent extinction.  We discuss several possible scenarios as the physical origins of the implied  SFR-$M_{\rm{star}}$ scaling law (\S\ref{origins}). We conclude that while we cannot completely rule out a possibility that UV-fainter galaxies have, on average, lower formation redshifts based on the observations, it is most likely that ``episodic star formation'' scenario is in best agreement with the available set of observations including clustering, abundance of low-mass galaxies, and population age constraints of high-redshift galaxies.  One interesting implication of such a scenario is that the star formation duty cycle increases with luminosity or halo mass. A direct result would be that galaxies with high SFR will continue to form stars with duty cycles close to unity and assemble their mass rapidly, while galaxies with low SFR will lag behind as their mass assembly is punctuated by relatively dormant phases. The differential evolution of galaxies at different luminosities presents a nontrivial test to theoretical models of galaxy formation. 

\acknowledgments
In memory of the late Ms. Michele Dufault, a very bright Yale undergraduate student who would have been a brilliant astrophysicist of the next generation. \\

We thank the anonymous referee for useful comments and suggestions. 
KSL gratefully acknowledges the generous support of Gilbert and Jaylee Mead for their namesake fellowship in the Yale Center for Astronomy and Astrophysics.  KSL also thanks Adam Muzzin, Danilo Marchesini, C. Meg Urry, Adriano Fontana, and Kevin Schawinski for useful discussions and suggestions. HM acknowledges support from Funda\c{c}\~{a}o para a Ci\^{e}ncia e a Tecnologia through the doctoral scholarship SFRH/BD/31338/2006 at Centro de Astronomia e Astrof\'{i}sica da Universidade de Lisboa (Portugal).

\bibliographystyle{/Users/kyoungsoolee/publications/apj}
\bibliography{/Users/kyoungsoolee/publications/apj-jour,/Users/kyoungsoolee/publications/myrefs}  

\end{document}